\documentclass[a4paper,11pt]{article}
\usepackage{graphicx}
\usepackage{amsmath,bbm,bm}
\usepackage{amssymb}
\usepackage[T1]{fontenc}
\usepackage{natbib}
\usepackage[margin=3cm]{geometry}
\usepackage{color}
\newtheorem{proposition}{Proposition}[section] 

\title{Estimating finite mixtures of semi-Markov chains: an application to the segmentation of temporal
sensory data}

\author{Herv\'{e} Cardot$^{(a)}$, Guillaume Lecuelle$^{(b)}$, Pascal Schlich$^{(b)}$ and Michel Visalli $^{(b)}$ \\
(a). Institut de Math\'ematiques de Bourgogne, UMR 5584 CNRS, \\ Universit\'{e} Bourgogne Franche-Comt\'{e}, F-21000 Dijon, France \\
(b). Centre des Sciences du Go\^{u}t et de l'Alimentation, \\ AgroSup Dijon, CNRS, INRA, \\
Universit\'{e} Bourgogne Franche-Comt\'{e}, F-21000 Dijon, France}

\begin{document}

\maketitle
\begin{abstract}
In food science, it is of great interest to get information about the temporal perception of aliments  to create new products, to modify  existing ones or more generally to understand the perception mechanisms. Temporal Dominance of Sensations (TDS) is a technique to measure temporal perception which consists in choosing sequentially attributes describing a food product over tasting. 
This work introduces new statistical models based on finite mixtures of semi-Markov chains  in order to describe data collected with the TDS protocol, allowing different temporal perceptions for a same product within a population. The identifiability of the parameters of such mixture models is discussed.  Sojourn time distributions are fitted with gamma probability distribution and a penalty is added to the log likelihood to ensure convergence of the EM algorithm to a non degenerate solution.  Information criterions are employed for determining the number of mixture components. Then, the individual qualitative trajectories are clustered  with the help of the maximum a posteriori probability (MAP) approach. A simulation study confirms the good behavior of the proposed estimation procedure. The methodology is illustrated on an example of consumers perception of a Gouda cheese and assesses the existence of several behaviors in terms of  perception of this product. 
\end{abstract}


\section{Introduction}

\begin{figure}
\centering
\makebox{\includegraphics[width=0.95 \textwidth]{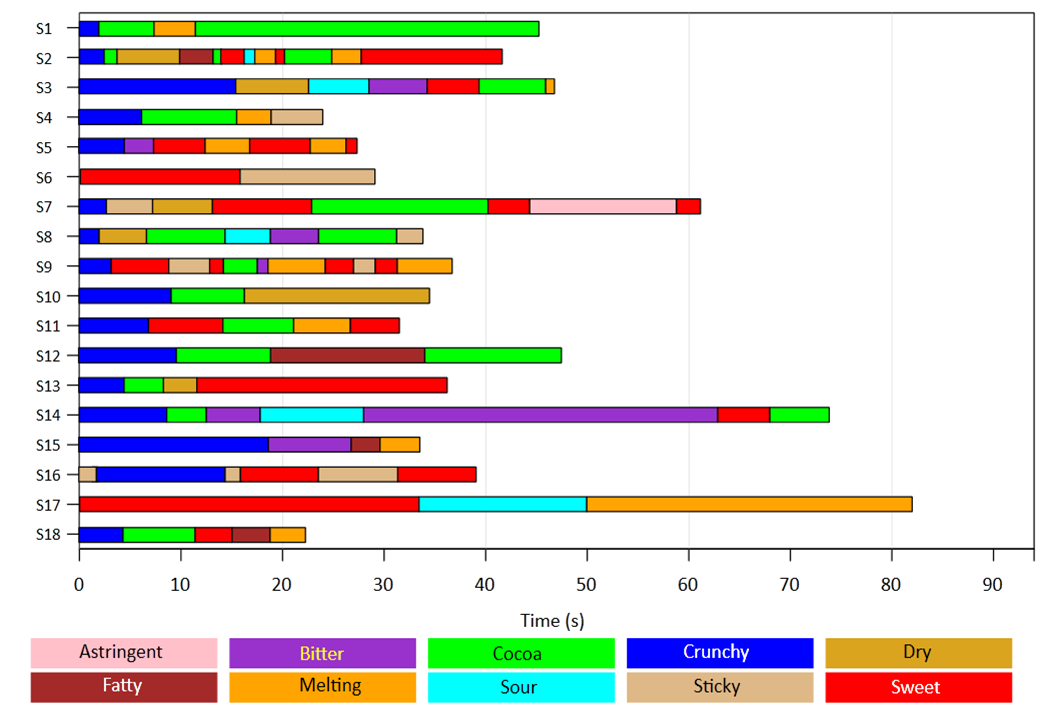}}
\caption{\label{fig01}Tasting of a chocolate with 70\% of cocoa by 18 panelists, denoted by $S_1$ to $S_{18}$, with 10 attributes. The bands represent the succession over time of the  dominant attributes selected by each panelist while tasting this chocolate. The figure has been obtained by means of the TimeSens$^\copyright$ software (www.timesens.com)}
\end{figure}

The development of food products is usually based on the measurement of product sensory perceptions from panels of consumers. Sensory perception while eating a food product has been acknowledged as a temporal process for 60 years \citep{Neilson1957}. Measuring the temporal sensory perception is a complex task and different approaches have been developed in sensory science (see \cite{Hort2017}). Recently a technique called Temporal Dominance of Sensations (TDS) has been introduced by \cite{Pineau2009}. A review on TDS can be found in \cite{Schlich2017}. The panelists have to describe the tasted product by choosing which attribute, among a list composed  of about ten items, corresponds to the most striking perception at a given time. This task results in sequences of attributes with choices and time of the choices. When an attribute is selected as dominant, it is considered as dominant until the panelist select another dominant attribute. At each time only one attribute can be dominant. An example of such an experiment for a chocolate tasting is presented in Figure~\ref{fig01} with data represented as bandplots. 

Some simple methods are currently used to describe such qualitative temporal data. Most of them rely on the observation of TDS curves, which consist in  representing the evolution along time of the proportions of the dominant attributes at a panel level. Even if this statistical approach can be very informative, such a tool only provides a mean panel overview and no information about the individual variability. Some quantitative analysis are used as complement \citep{galmarini} but these methods only consider dominance durations (the time spent as dominant for each attribute). None of these approaches takes into account the whole complexity of TDS data: choices of dominant attribute, order of the choices and dominance durations that are sojourn times in the successive dominant attributes. Recently, \cite{franczak} proposed to model TDS data with Markov chains. The Markov hypothesis, meaning that the probability of the next choice of dominant attribute only depends on the current dominant attribute, seems to be reasonable from a sensory perspective. However, the Markov hypothesis imposes strong restrictions on the sojourn time distribution which should be geometrically distributed when considering a discrete time process, or exponentially distributed when considering a continuous time process (see {\it e.g.} \cite{MR1600720} for a general presentation of Markov chains).
In a recent paper by \cite{Lecuelle} it has been noted that the sojourn time distributions  were not distributed according to a geometric law.  Consequently, it has been  proposed  to model TDS data with  semi-Markov chains (SMC)  and it has been shown that  allowing arbitrarily distributed sojourn times permits to get a better fit to the data.  Note that  approaches based on multivariate categorical data are not adapted for TDS data since we  observe sequences with a random number of visited states (see Figure~\ref{fig01}).  SMC, or Markov renewal processes, which have been introduced more than sixty years ago \citep{Levy, Smith}, are now widely used in numerous fields of science such as queuing theory, reliability and maintenance, survival analysis, performance evaluation, biology, DNA analysis, risk processes, insurance and finance or earthquake modeling (see {\it e.g.} \cite{Barbu2008} and references therein). 

It has often been suggested by sensory scientists \citep{Jaeger2017} that consumers form non homogeneous populations and heterogeneity in consumers' food products perception has been established  in \cite{Prutkin2000}.  To take into account heterogeneity among individuals and avoid conclusions on a non-existing "average consumer", consumer segmentation is a recommended strategy (\cite{Koster2009}, \cite{Meiselman2013}). Introducing  mixtures for modeling the different perceptions of a sample of panelists for a same product can be of real interest.

A mixture model \citep{Mclachlan2000, Melnykov2010} is a probabilistic model enabling to represent the presence of sub populations within an overall population. Finite mixture models are widely used in numerous fields of science such as biology or economy  because they offer probabilistic tools for performing clustering.   Mixture models are commonly used with the Gaussian distribution  but they can also be used with any parametric model (see the numerous examples in \cite{Fruhwirth2006} as well as  \cite{Banfield} or \cite{McNicholas2016}). For temporal data, mixtures  of Markov chains have been used in different fields such as finance \citep{Frydman}, computer science \citep{Song}, road traffic estimation \citep{Lawlor} or labor economy \citep{MR2719656}. In continuous time and continuous response, \cite{Delattre}  introduce mixtures of stochastic differential equations and use a classification rule based on estimated posterior probabilities to cluster growth curves.  However, as far as we know, the present work is the first one that considers mixtures of semi-Markov processes. 
The purpose of this article is to estimate mixtures of Semi-Markov chains, in discrete or continuous time,  to perform a segmentation of a sample of panelists into groups with similar perceptions. The methodology developed in this article can be useful in many domains for which the aim is to analyze  and perform a segmentation of panels of categorical trajectories.

Identifiability is a crucial issue for mixture models (see \cite{titterington1985statistical} and \cite{Fruhwirth2006}) and we show under general conditions that,  when  identifiable parametric models are considered for the distribution of sojourn times, the parameters of the model are identifiable up to label swapping. The estimation of the parameters is performed with the EM algorithm \citep{McLachlan2008} in which a penalty may be added to avoid degenerate solutions. In our sensory analysis example, sojourn times are fitted with gamma distributions and as explained in \cite{MR3578958}, the likelihood is generally unbounded in case of mixtures of  gamma distributions. We thus consider a penalized likelihood criterion that leads to  more stable estimates and permits to avoid degenerate solutions. 
The number of mixture components being generally unknown, an information criterion is employed to select the number of sub populations that should be considered (see \cite{MR2719656} for a discussion about model selection in the context of mixtures of Markov chains). Then, the observed trajectories can be clustered thanks to the maximum a posteriori probability (MAP) classification approach (see \cite{Fruhwirth2006}).

The proposed method is illustrated on a dataset from the European Sensory Network \citep{Thomas2017}. This dataset includes TDS data for 4 Gouda cheeses tasted by 665 consumers according to 10 attributes. A mixture of SMC with gamma sojourn time distributions is adjusted to fit the data.  

The article is organized as follows. Section 2 presents the mixture models and discusses the identifiability issue. Section 3 presents the EM algorithm employed for the estimation of the parameters of the mixture, the proportions and the number of components. Section 4 evaluates the performances of the statistical methods through a simulation study and Section 5 provides an illustration of the proposed method on cheese tasting data. Concluding remarks and discussion are given in Section~6.

\section{Stochastic model and notations}

\subsection{Markov renewal processes and finite mixtures of Markov renewal processes}

\begin{figure}
\centering
\makebox{\includegraphics[width=0.8 \textwidth]{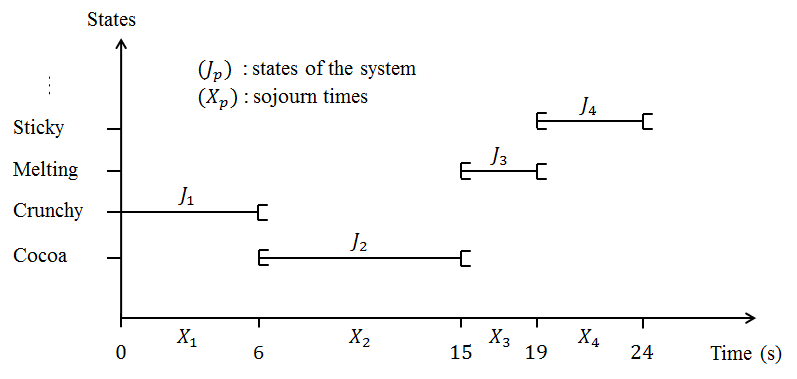}}
\caption{\label{fig_SMP}Modeling of sequence $S4$ (see Figure \ref{fig01}) with a Markov renewal process $(J_p,X_p)_{p\geq 1}$. The successive states chosen by the panelist are $J_1=$Crunchy, $J_2=$Cocoa, $J_3=$Melting and $J_4=$Sticky}
\end{figure}

Consider a finite state homogeneous Markov chain $(J_p)_{p\geq 1}$, taking values in the finite state space $\mathcal{S} = \{1, \ldots, D\}$, with transition matrix $\mathbf{P}$, whose  generic elements are $P_{\ell j} = \Pr \left[ J_{p+1}=j | J_p=\ell\right]$, $\ell, j \in \mathcal{S}$.  Consider the random sequence $(X_p)_{p \geq 1}$ made by the successive sojourn times in the visited states. For each $p \geq 1$, $X_p$ represents the sojourn time at state $J_p$ and takes values in $T={1,2,\ldots}$ if time, denoted by $t$, is discrete and in $T = [0,+\infty[$ if time is continuous. For $j \neq \ell$, we denote by $\Phi_{\ell j}(t) = \Pr \left[X_p \leq t \ | \ J_p=\ell, J_{p+1}=j \right]$, the cumulative distribution function of the sojourn time given the current and the next  states of the random process $(J_p)_{p \geq 1}$.
We suppose that the random process $(J_p,X_p)_{p\geq 1}$ satisfies the Markov property, for all $t \in T$, $\ell \in \mathcal{S}$ and $j \neq \ell$,  
\begin{align}
\Pr \left[ J_{p+1} =j, X_p \leq t \ | \ J_p=\ell, J_{p-1}, \cdots , J_1, X_{p-1}, \cdots, X_{1}\right] 
%= \mbox{Pr}\left[ J_{n+1}=j, X_n \leq t \ | \ J_n=\ell \right] \nonumber \\
&= P_{\ell j} \ \Phi_{\ell j}(t). 
\end{align}
The process $(J_p,X_p)_{p\geq 1}$ is called a Markov renewal process, whereas the stochastic process giving the state of the system at every time $t \in T$ is called a semi-Markov process (see {\it e.g.} \cite{Pyke1961} or \cite{Barbu2008}). 
For identifiability reasons,  it is also supposed that $P_{j j} = 0$, for all $j \in \mathcal{S}$, so that at each jump, the system cannot remain in the same state.  
To avoid trajectories with an infinite number of visited states, we also suppose that the semi-Markov chain is regular (see \cite{Pyke1961}). This is true for gamma distributed sojourn times considered in the application, and more generally under the very weak condition that the cumulative distribution function is continuous at 0 with $\lim_{t\to 0_+}\Phi_{\ell j}(t) = 0$. Finally, to completely characterize the law of $(J_p,X_p)_{p\geq 1}$ we define the vector $\boldsymbol{\alpha} =( \alpha_1, \ldots, \alpha_D)$ of initialization probabilities
\begin{align}
\alpha_j &= \mbox{Pr} \left[ J_1 =  j \right], \quad j \in \mathcal{S}.
\end{align}
The example given in Figure~\ref{fig_SMP} describes the representation, in terms of semi-Markov trajectory, of the $4^{th}$ TDS sequence of the dataset presented in Figure~\ref{fig01}.

The distribution of the semi-Markov process $(J_p,X_p)_{p \geq 1}$ is completely characterized by the set of parameters $(\boldsymbol{\alpha}, \mathbf{P}, \Phi_{\ell j}, \ell, j \neq \ell \in \mathcal{S})$ and in the following its probability law is denoted by $\mbox{Law} \left(\boldsymbol{\alpha}, \mathbf{P}, \Phi_{\ell j}, \ell, j \neq \ell \in \mathcal{S} \right)$.
 
Let us consider now  $G$ independent semi-Markov processes taking values in the same state space $\mathcal{S}$, and for $g=1, \ldots, G$, the initialization vector of probabilities $\boldsymbol{\alpha}^g$, the transition matrix $\mathbf{P}^g$,  and the cumulative distribution functions for the sojourn times $\Phi_{\ell j}^g(t), \ t \in T$. Denoting by $\pi_g>0$, the probability of observing a Markov renewal process with parameters  $\left(\boldsymbol{\alpha}^g, \mathbf{P}^g, \Phi_{\ell j}^g, \ell, j \neq \ell \in \mathcal{S} \right)$, we consider the finite mixture process $(J_p^\pi,X_p^\pi)_{p\geq 1}$ whose law is given by
\begin{align}
\sum_{g=1}^G \pi_g \mbox{Law}\left(\boldsymbol{\alpha}^g, \mathbf{P}^g, \Phi_{\ell j}^g, \ell, j \neq \ell \in \mathcal{S} \right).
\label{def:mixtureMRP}
\end{align}
The following proposition states that a finite mixture of Markov renewal processes is a Markov renewal process.
\begin{proposition}
The process $(J_p^\pi,X_p^\pi)_{p\geq 1}$ is a Markov renewal process with parameters 
\[
\left( \sum_{g=1}^G \pi_g \boldsymbol{\alpha}^g, \sum_{g=1}^G \pi_g \mathbf{P}^g, \sum_{g=1}^G \pi_g \Phi_{\ell j}^g, \ell, j \neq \ell \in \mathcal{S} \right).
\]
\label{prop:mixtureMRP}
\end{proposition}

\subsection{The identifiability issue}

Identifiability of mixture models can be a complicated issue (see {\it e.g.} \cite{Teicher63}, \cite{YakowitzSpragins1968},  \cite{titterington1985statistical} or \cite{AMR2009}). %For instance it is now well known that mixtures of binomial distribution are identifiable if and only if $G \leq (n+1)/2$. 
However, identifiability of the parameters of a stochastic model is a very important condition to ensure the convergence of estimation algorithms to a unique value.  We consider here a parametric framework and we are interested in models defined by a family of distributions $\mathcal{F}(\boldsymbol{\Theta}) = \{ \mbox{Law}(\boldsymbol{\theta}), \boldsymbol{\theta} \in \boldsymbol{\Theta} \}$ where $\boldsymbol{\Theta} \subset \mathbb{R}^q$ is the parameter space and $\boldsymbol{\theta}$ is a vector of parameters characterizing the probability distribution. We consider convex combinations of  probability laws in   $\mathcal{F}(\boldsymbol{\Theta})$, $\sum_{g=1}^G c_g \mbox{Law}(\boldsymbol{\theta}_g)$, with $\sum_g {c_g}=1$,  $c_g>0$ and $\boldsymbol{\theta}_g \in \boldsymbol{\Theta}$, for $g=1, \ldots, G$.

Adopting the same definition as in \cite{YakowitzSpragins1968}, we say that the finite mixtures are identifiable in the family $\mathcal{F}(\boldsymbol{\Theta})$ if and only if the convex hull of $\mathcal{F}(\boldsymbol{\Theta})$ has the uniqueness representation property:
\begin{align}
\sum_{g=1}^G c_g  \mbox{Law}(\boldsymbol{\theta}_g) & = \sum_{h=1}^H c'_h \mbox{Law}(\boldsymbol{\theta}_h)
\label{identif:finitemix}
\end{align}
implies $G=H$ and for each $g \in {1, \ldots, G}$ there is some $h \in \{1, \ldots, G\}$ such that $c_g = c'_h$ and $\boldsymbol{\theta}_g = \boldsymbol{\theta}_h$.

Moreover, it has been proven in \cite{YakowitzSpragins1968} that finite mixtures of a family $\mathcal{F}(\boldsymbol{\Theta})$ are identifiable if and only if the cumulative distribution functions of the  elements are linearly independent. 

We suppose from now that the family of distributions of the sojourn times is parametric, $\Phi_{\ell j} (t) = \Phi(t, \boldsymbol{\Gamma}_{\ell j})$, with $\boldsymbol{\Gamma}_{\ell j} \in \mathbb{R}^d$.  Classical parametric distributions of sojourn times are the negative binomial distribution if time is discrete ($d=2$), and exponential ($d=1$) or gamma distributions ($d=2$) if time is continuous. In our renewal Markov processes framework, a parameter $\boldsymbol{\theta}_g$ will be of the form $\boldsymbol{\theta}_g = \left(\boldsymbol{\alpha}^g,  \mathbf{P}^g, \boldsymbol{\Gamma}^g_{\ell j}, \ell \in \mathcal{S}, j \neq \ell \in \mathcal{S}\right)$.
%Taking account of the identifiability constraints on the initialization probabilities and the transition probabilities, the number of parameters require to characterize the law of each component of the mixture is equal to $D-1 + D(D-2) + D(D-1)d$.  
It is shown in \cite{Teicher63} that finite mixtures of Gamma distributions are identifiable whereas it is proven in \cite{YakowitzSpragins1968} that finite mixtures of exponential distributions as well as negative binomial distributions are also identifiable. More recently, it has been shown in \cite{gupta2016mixtures}, under technical assumptions, that almost all  finite mixtures of Markov chains with $D$ states are identifiable provided that at least two consecutive transitions can be observed and the number of mixture components is not too large compared to the number of states,  more precisely $D \geq 2G$.

As shown in the next proposition, the identifiability of mixtures of renewal Markov processes can be assessed under weaker conditions than those required for mixtures of Markov chains because the sojourn time distributions are not directly related to transition probabilities and there is no need to introduce any particular condition on the number of mixture components or on the number of states of the Markov chain. See also \cite{GassiatCleynenRobin} for an intermediate identifiability result stated for Hidden Markov Chains, which does not impose any condition on the number of states and mixture components but is based on the knowledge of the law of at least two consecutive transitions.

For sake of simplicity we assume that all the initialization probabilities $\alpha_\ell^g$ and all the transition probabilities $P_{\ell j}^g$ are strictly positive. This ensures that all the sojourn time distributions can be observed by considering the law of  $(J_1^\pi, X_1^\pi,J_2^\pi)$.

\begin{itemize}
\item[$\mathbf{(H1)}$] $\displaystyle \forall g \in \{1, \ldots, G\}, \forall \ell \in \mathcal{S}, \  \alpha_\ell^g >0 \ \mbox{ and } \ \forall j \neq \ell, \ P^g_{\ell j} >0.$
\end{itemize}

We also need to add the following  hypothesis which means that two subpopulations $g$ and $g'$ cannot have exactly the same set of parameters for the distributions of duration times.
\begin{itemize}
\item[$\mathbf{(H2)}$] $\displaystyle \forall g \in \{1, \ldots, G\} \mbox{ and } \forall g' \neq g,  \exists \ell \in \mathcal{S} \mbox{ and } j \neq \ell \mbox{ such that }   \boldsymbol{\Gamma}^g_{\ell j} \neq  \boldsymbol{\Gamma}^{g'}_{\ell j}.$
\end{itemize}
 If this condition  is not fulfilled, we may have two mixture components whose  sojourn time distributions are exactly the same. In that case, we are not able  to distinguish the two corresponding subpopulations according to their sojourn times.

\begin{proposition}\label{prop:2.2} 
Suppose that the  family of sojourn time distributions is identifiable and that hypotheses $\mathbf{(H1)}$ and $\mathbf{(H2)}$ hold. Then all the finite mixtures  from the family $\mathcal{F}(\boldsymbol{\Theta})$ can be identified when  the law of the sequence $(J_1^\pi, X_1^\pi,J_2^\pi)$ drawn form  a mixture of renewal Markov processes is known.
\label{prop:identif}
\end{proposition}

In other words, it is possible to identify the parameters of a finite mixture from $\mathcal{F}(\boldsymbol{\Theta})$ provided that we can observe at least one transition and the first sojourn times and the first state. Note that the condition $\boldsymbol{\alpha}^g_\ell >0$, which is also required in  \cite{gupta2016mixtures}, ensures that all the possible transitions can be observed during the first transition. This hypothesis could be weakened by considering the law of mixture sequences with more than one transition. The condition on the transition probabilities that must be strictly positive is essentially of technical nature and allows to simplify the demonstration. Note that $P_{\ell j}^g=0$ means that the transition from $\ell$ to $j$ is never observed so that we cannot associate a duration time distribution to the transition from state $\ell$ to state $j$ in mixture component $g$.   
Without assumption $\mathbf{(H1)}$, we should restrict the set of indices related to the sojourn times to the set  corresponding to strictly positive transition probabilities.

\section{Maximum likelihood estimation and model selection}\label{sec:MLE}

Suppose we have  a sample of $n$ independent consumers, for which we may consider $B$ independent and identically distributed replications of the tasting experiment. For each consumer $i$, with $i=1, 2, \ldots, n$, we thus get  $B$ sequences $S_i^b$, for $b=1, \ldots, B$,  observed for $t \leq T_i^b$ and denoted by,
\begin{align}
S_i^b &= (J_1^{i,b}, X_1^{i,b}, \dots, J_{N(T_i^b)-1}^{i,b},X_{N(T_i^b)-1}^{i,b},J_{N(T_i^b)}^{i,b},X_{N(T_i^b)}^{i,b}),  
\label{def:seqb}
\end{align} 
where $N(T_i^b)$ is the random number of visited states by consumer $i$ during replication $b$. We suppose that $N(T_i^b) \geq 2$.

We suppose that the observed trajectories $S_1^1, \ldots, S_1^B, \ldots, S_n^1, \ldots, S_n^B$ are drawn from  a mixture of $G$ semi-Markov processes whose law is given in (\ref{def:mixtureMRP}) and we aim at estimating the parameters which characterize the law of the mixture: the vector of mixture proportions $\boldsymbol{\pi} = (\pi_1, \ldots, \pi_G)$, and $(\boldsymbol{\alpha}^g, \mathbf{P}^g, \Phi_{\ell j}^g, \ell \in \mathcal{S}, j \neq \ell \in \mathcal{S})$, for $g=1, \ldots, G$ which characterize the law of the semi-Markov processes for each mixture component. We suppose in this Section that the number $G$ of components is known.

\subsection{The particular case of gamma distributed sojourn times with replications and no anticipation}\label{sec:SMCGamma}

In our sensory examples, sojourn times are positive and continuous random variables and suppose that they are distributed according to gamma distributions. The choice of the gamma distribution is motivated by its simplicity  and its ability to fit  sojourn time distributions with many different shapes. The density depends on two parameters, the shape parameter $a>0$ and $\lambda>0$, and is defined as follows,
\begin{align*}
f(t,a,\lambda) &= \frac{t^{a-1} \lambda^{a} \exp(-\lambda t)}{\Gamma(a)},  \quad t \geq 0, 
\end{align*}
where $\Gamma(a)$ is the gamma function.  The corresponding expected value is $a/\lambda$ and the variance $a/\lambda^2$.

We suppose, as in  \cite{Lecuelle},  that the sojourn time distribution only depends on the current state,
\begin{align}
\Pr[X^\pi_p \leq t |  J_p = \ell, J_{p+1} = j, Z = g] &= \Pr[X^g_1 \leq t |  J_1 = \ell] 
\label{def:sejoursimple}
\end{align}
so that there is no anticipation, in some sense, of the next dominant attribute. 
This assumption, which seems relevant in a food tasting context, also allows us to deal with moderate size samples by reducing significantly the number of parameters to be estimated. In that case, hypothesis $(\textbf{H2})$ means that for each mixture components $g$ and $g'$, there is at least one state $\ell$ such that the two cumulative distributions $\Pr[X^g_1 \leq t |  J_1 = \ell]$ and $\Pr[X^{g'}_1 \leq t |  J_1 = \ell]$ are not equal.  If we denote by $d$ the number of parameters required to characterize each sojourn time distribution,  we only need, with this simplification, to estimate $GDd$ parameters to characterize the sojourn time distributions instead of $GD(D-1)d$ in the more general setting studied in previous Section. Note that form now on $d=2$, which corresponds to the particular case of gamma distributed sojourn times. 

\subsection{The likelihood}

By successive conditioning, the likelihood related to a statistical unit $i$ with $B$ independent replications drawn from a Markov renewal process with parameters $\boldsymbol{\theta}_g = \left(\boldsymbol{\alpha}^g, \mathbf{P}^g, (a_{\ell g}, \lambda_{\ell g}), \ell \in \mathcal{S} \right)$ can be written
\begin{align}
L_g(S_i^1, \ldots, S_i^B ; \boldsymbol{\theta}_g) &= \prod_{b=1}^B L_g(S_i^b ; \boldsymbol{\theta}_g) \nonumber \\
 &= \prod_{b=1}^B \left[ \alpha _ {J_1^{i,b}}^g \phi^g_{J_1^{i,b}}(X_{1}^{i,b}) \prod_{k=2}^{N(T_i^b)}  \mathbf{P}^g_{J_{k-1}^{i,b} J_k^{i,b}} \phi^g_{J_k^{i,b}}(X_{k}^{i,b}) \right] 
\label{def:likSib}
\end{align}
where $\phi^g_{\ell}(x) = \int_0^x f(t,a_{\ell g},\lambda_{\ell g}) dt$ is the cumulative distribution function for a gamma random variable with parameters $a=a_{\ell g}$ and  $\lambda=\lambda_{\ell g}$,

If we do not suppose anymore that the mixture component from which unit $i$ arises is known, the log likelihood under the mixture model of the $nB$ trajectories becomes

\begin{align}
\ln  L(S_1^1, \ldots, S_n^B ; \boldsymbol{\theta})  &= \sum_{i=1}^n \ln \left(\sum_{g=1}^G \pi_g \prod_{b=1}^B L_g(S_i^b ; \boldsymbol{\theta}_g) \right),
 \label{def:loglikS1Sn}
 \end{align}
where $\boldsymbol{\theta}=(\boldsymbol{\pi}, \boldsymbol{\theta}_1, \ldots, \boldsymbol{\theta}_G)$ is the set of  parameters of the mixture model.
A direct maximization of the log-likelihood (\ref{def:loglikS1Sn}), according to  $\boldsymbol{\theta}$ is cumbersome and classical optimization algorithm are generally not suitable to deal with that kind of problem (see {\it e.g} \cite{McLachlan2008}). The EM algorithm, presented below, is preferred because it allows the optimization procedure to be decomposed into two simple steps. 

\subsection{The EM algorithm}

The Expectation Maximization (EM)  algorithm is a very useful algorithm that has first been designed to perform maximum likelihood estimation for incomplete data problems (see  \cite{Dempster1977}). It is an iterative optimization technique of the likelihood that can be very effective for estimating mixture models by considering the unknown mixture components as missing observations (see \cite{Mclachlan2000}). 

Let us introduce the missing mixture component indicators, $Z_i$, for $i=1, \ldots, n$, which are vectors with $G$ elements, composed of 1 one and $G-1$ zeros and that indicates from which component of the mixture the trajectory $S_i$ arises. In other words,  if $S_i$ has been generated by the $g^{th}$ mixture component then $Z_{ig}=1$ and $Z_{i\ell}=0$ for $\ell \neq g$. The complete data log-likelihood can be written as follows:
\begin{align}
\ln  L_c( S_1^1, \ldots, S_1^B, Z_1, \ldots, S_n^1, \ldots, S_n^B, Z_n; \boldsymbol{\theta} ) & = \sum_{i=1}^n \sum_{g=1}^G Z_{ig} \ln \left( \pi _g \prod_{b=1}^B L_g(S_i^b ; \boldsymbol{\theta}_g) \right) \nonumber\\
 &= \sum_{i=1}^n \sum_{g=1}^G Z_{ig} \ln \pi _g  + \sum_{i=1}^n \sum_{g=1}^G Z_{ig} \sum_{b=1}^B \ln L_g(S_i^b; \boldsymbol{\theta}_g). 
\label{completedLL}
\end{align}
This function is  much easier to maximize, according to $\boldsymbol{\theta}$, than the log-likelihood function given in (\ref{def:loglikS1Sn}).

An initial value $\boldsymbol{\theta}^{(0)}$ of the parameters must be carefully chosen before starting the algorithm. The choice of the starting point can be of great importance and is discussed in Section~\ref{sec:startpoint}.   
The EM algorithm proceeds iteratively according to the following scheme.  
%The EM algorithm is stopped after a fixed number of iterations chosen large enough so that the sequence $\boldsymbol{\theta}^{(m)}$ is stabilized. 
Suppose an estimate of $\boldsymbol{\theta}$, denoted by $\boldsymbol{\theta}^{(m-1)}$, has been calculated at step $m-1$, with $m\geq 1$.

\subsubsection*{E-step}
The expectation (E) step consists in computing  the expected log-likelihood of the complete data given the observed trajectories and the value of the parameters estimated during the previous iteration. We define
\begin{align}
Q(\boldsymbol{\theta},\boldsymbol{\theta}^{(m-1)}) &= \mbox{E} \left[ \ln  L_c \left( S_1, Z_1, \ldots, S_n, Z_n; \boldsymbol{\theta} \right) | S_1, \ldots, S_n,  \boldsymbol{\theta}^{(m-1)} \right] \nonumber \\
 &= \sum_{i=1}^n \sum_{g=1}^G \widehat{Z}_{ig}^{(m)} \sum_{b=1}^B \ln L_g(S_i^b; \boldsymbol{\theta}_g^{(m-1)} ) + \sum_{i=1}^n \sum_{g=1}^G \widehat{Z}_{ig}^{(m)} \ln \pi _g^{(m-1)}, 
\label{def:Q} 
\end{align} 
with $\widehat{Z}_{ig}^{(m)} = \mbox{E}[Z_{ig} | S_1, \ldots, S_n, \boldsymbol{\theta}^{(m-1)}]$,
the conditional probability for $S_i$ to be generated by the component $g$ of a mixture model with parameters $\boldsymbol{\theta}^{(m-1)}$, where $\boldsymbol{\theta}^{(m-1)}$ is the value of the set of parameters computed  at  previous iteration.
We get with Bayes theorem, 
%the following best approximation of the indicator random variables $Z_{ig}$ given the data and the estimated value of the parameter $\boldsymbol{\theta}$ obtained at past iteration. 
\begin{align}
\widehat{Z}_{ig}^{(m)}&=\Pr(Z_{ig}=1 | S_i; \boldsymbol{\theta}^{(m-1)}) \nonumber \\
&= \frac{ \displaystyle \pi_g^{(m-1)} \prod_{b=1}^B L_g(S_i^b ; \boldsymbol{\theta}^{(m-1)})}{\displaystyle \sum_{j=1}^G \pi_j^{(m-1)} \prod_{b=1}^B L_j(S_i^b ; \boldsymbol{\theta}^{(m-1)})}.
\end{align}

\subsubsection*{M-step}
 The maximization (M) step consists in updating the value of parameter $\boldsymbol{\theta}$ given  the expected values of  $\hat{Z}_{ig}$, for $g=1, \ldots, G$ and $i=1, \ldots, n$, by looking for the maximum, according to $\boldsymbol{\theta}$, of  the function $Q(\boldsymbol{\theta},\boldsymbol{\theta}^{(m-1)})$  defined in (\ref{def:Q}). 
The mixture probabilities $\pi_g$ only appear in the second term at the righthand side of (\ref{def:Q}). The new estimates  at step $m$  are obtained by solving
\begin{align}
\frac{\partial}{\partial \pi_g} \left[ \sum_{i=1}^n \sum_{g=1}^G \hat{Z}_{ig}^{(m)} \ln \left( \pi _g \right) + \lambda \left(\sum_{g=1}^G \pi _g -1\right)\right]=0,
\end{align}
where $\lambda$ is the Lagrange multiplier associated to the constraint $\sum_{g=1}^G \pi _g=1$.
We get the standard solution, $\pi _g^{(m)} = n^{-1} n_g^{(m)}$,
with $n_g^{(m)}=\sum_{i=1}^n \widehat{Z}_{ig}^{(m)}$.

The $G$ Markov chain transition matrices and initialization probabilities  $(\boldsymbol{\alpha}_g, \mathbf{P}_g, g=1, \ldots, G)$ as well as the parameters related to the sojourn time distributions  $\left(a_{\ell g}, \lambda_{\ell g} \right)$ % , \ell \in \mathcal{S}, g =1, \ldots, G \right)$ 
 are updated by maximizing the first term at the right-hand side of (\ref{def:Q}).

Thanks to the multiplicative structure of the likelihood (\ref{def:likSib}) given the mixture component,  the first term at the righthand side of (\ref{def:Q}) can be written as the sum of two distinct functions, where the first one only depends on the semi-Markov chains parameters $(\boldsymbol{\alpha}_g, \mathbf{P}_g, g=1, \ldots, G)$ whereas the second one only depends on the sojourn time distributions $( a_{\ell g}, \lambda_{\ell g} , \ell \in \mathcal{S}, g =1, \ldots, G)$. Thus, these two sets of parameters can be estimated separately by maximizing each part of the log-likelihood during the M-step. 
%with the weights $\hat{Z}_{ig}^{(m)}$ on individuals. 
Introducing again Lagrange multipliers, this yields the standard solution for the transition probabilities estimators as well as the initialization probabilities:
\begin{align}
\widehat{\alpha}_j^{g(m)} & = \frac{\sum_{i=1}^n \widehat{Z}_{ig}^{(m)} \sum_{b=1}^B \mathbbm{1}_{\{J_1^{i,b} = j\}}}{B \sum_{i=1}^n \widehat{Z}_{ig}^{(m)}}, \qquad \widehat{\mathbf{P}}^{g(m)}_{hj} 
= \frac{\sum_{i=1}^n \widehat{Z}_{ig}^{(m)} \sum_{b=1}^Bn_{hj}^{ib}}{\sum_{\ell=1}^D \sum_{i=1}^n \widehat{Z}_{ig}^{(m)}\sum_{b=1}^B n_{h \ell}^{ib}}, 
\end{align}
where $n_{hj}^{ib}$ is the number of $h\rightarrow j$ transitions for trajectory  $S_i^b$.

%We now derive the EM algorithm in the particular case of Gamma distributed sojourn times, with parameters only depending on the current state. 
It is  shown in \cite{MR3578958} that for gamma mixture models the log likelihood is not bounded.  Intuitively, the degeneracy  comes from the fact that  if the ratio $a_{\ell g}/\lambda_{\ell g}$, which corresponds to the expected sojourn time in state $\ell$ for mixture $g$ is kept constant, while $a_{\ell g}$ is tending to infinity, then the corresponding variance $a_{\ell g}/\lambda^2_{\ell g}$ will tend to zero and the corresponding gamma density, mimicking the Dirac distribution at $a_{\ell g}/\lambda_{\ell g}$, will not be bounded.   
Consequently, to avoid such degenerate solution, it may be preferable to introduce a penalization in the M-step that prevents the parameters $a_{\ell g}$ from becoming too large. Thus, we add to function $Q$, defined in (\ref{def:Q}), a  penalty similar to the penalty given in \cite{MR3578958} and defined as follows
\begin{align}
P_{en} \left(a_{\ell g}, \ell \in \mathcal{S}, g=1, \ldots, G \right) &= - \frac{1}{\sqrt{\sum_{i=1}^n \sum_{b=1}^B N\left(T_i^b\right)}} \sum_{g=1}^G \sum_{\ell \in \mathcal{S}} \left( a_{\ell g} + \ln a_{ \ell g} \right).
%\frac{-1}{\sqrt{n}} \sum_{g=1}^G \sum_{\ell \in \mathcal{S}} \left( a_{\ell g} + \ln a_{ \ell g} \right)  
\label{def:penalty}
\end{align}
Note that this penalty does not need to take into account the parameters $\lambda_{\ell g}$ of the gamma distributions. Its effect decreases as the sample size and the number of observed transitions increase.

Finally, the parameters of the sojourn time distributions can be estimated by maximizing the following expected partial penalized log-likelihood:
\begin{align}
 \sum_{i=1}^n \sum_{g=1}^G \widehat{Z}_{ig}^{(m)} \sum_{b=1}^B \sum_{k=1}^{N(T_{i}^b)} \ln \phi_{J_k^{i,b}}^g \left(X_{k}^{i,b} \right) +P_{en}\left(a_{\ell g}, \ell \in \mathcal{S}, g=1, \ldots, G\right),
   %- \frac{1}{\sqrt{\sum_{i=1}^n \sum_{b=1}^B N\left(T_i^b\right)}} \sum_{g=1}^G \sum_{\ell \in \mathcal{S}} \left( a_{\ell g} + \ln a_{ \ell g} \right).
\label{def:mlepen} 
\end{align}
with classical optimization procedures.

\medskip

Once the algorithm has converged, model-based clustering of the  observed sequences is  performed by considering the maximum \textit{a posteriori}  probability (MAP) criterion, defined as follows: $\mbox{MAP}(\hat{Z}_{ig})=1$ if $g= \mbox{arg} \max_h(\hat{Z}_{ih})$ and $\mbox{MAP}(\hat{Z}_{ig})=0$ otherwise.

\subsection{Choosing the starting point of the EM algorithm}\label{sec:startpoint} 
A crucial issue for the EM algorithm is the choice of the value of the starting point $\boldsymbol{\theta}^{(0)}$. 
It is shown in \cite{galmarini} that the time spent in each state provides interesting indicator to study TDS data. Thus, we have chosen to select the initial values of the EM algorithm by considering the Hartigan-Wong k-means algorithm \citep{hartigan1979} applied to the $D$ dimensional vector of mean sojourn times in each state, with the Euclidean distance and $k=G$ clusters.
A heuristic justification can be given by the fact that the identification of the mixture components seems to be easier for sojourn times. Indeed as seen in the proof of Proposition~\ref{prop:2.2}, the sojourn time distribution of any finite mixture of Markov renewal processes is identifiable when the family of sojourn time distributions is identifiable. Then, the method of moments is employed to get the initial values for the transition matrices and for the parameters of the gamma distributions.

\subsection{Selection of the number of mixture components}
\label{sec:BIC}

When the number of components $G$ is unknown, an information criterion can be used  to select the number of mixture components (see \cite{Mclachlan2000} for a detailed presentation of the various approaches developed in the literature). Such information criteria rely on a compromise between the fit to the data and the complexity of the considered model, more complex models being less desirable. The Bayes Information Criterion (BIC), which has good asymptotic properties  (see \cite{Keribin2000}), is simple to compute and seems effective  to select the number of components. We choose to  define the BIC criterion as follows,
\begin{align}
\mbox{BIC}(G) &= q \ln \left(nB\right)  - 2 \ln L \left(S_1^b, \ldots, S_n^b, b=1,\ldots, B; \ \hat{\boldsymbol{\theta}} (G) \right),
\label{def:BIC}
\end{align}
where $\hat{\boldsymbol{\theta}}(G)$ is the estimation of $\boldsymbol{\theta}$ when a mixture of $G$ components has been considered  and $q = q(\boldsymbol{\theta}(G))$ is the number of free parameters to be estimated. 
Note that $nB$ corresponds to the number of independent observations in classical mixture models and could be different in our setting of temporal data. Indeed, it is not completely clear which value should be considered for  the sample size since, for each trajectory, we have observations that are correlated over time (see \cite{MR2719656} for a discussion in a similar context of mixtures of Markov chains) and we could also take account of the number of observed transitions. In the particular setting described in Section~\ref{sec:SMCGamma} and taking into account the fact that $P_{\ell \ell}^g=0$ for all $g=1, \ldots, G$ and all $\ell \in \mathcal{S}$, we get
$q=G-1 + G(D-1 + D(D-2) +  Dd) = GD(D+d-1) - 1$, with $d=2$ for the two parameters gamma distribution. If there is one absorbing state in $\mathcal{S}$, as in the example in Section~\ref{sec:gouda}, and we suppose that it is not possible that the first observed state is this absorbing state, then $q= G-1 + G(D-2 + (D-1)(D-2) + (D-1)d)$.

Other popular criteria are the Akaike information criterion (AIC), in which the term  $q \ln nB$  in (\ref{def:BIC}) which penalize the complexity of the model is replaced by $2 q$ and the corrected AIC, denoted by $\mbox{AIC}_c$ in which the term $q \ln nB$ is replaced by $2q + \frac{2q(q+1)}{nB-q-1}$.

\section{A simulation study}

A simulation study is conducted to evaluate the performances of the penalized and unpenalized EM algorithms under various mixture scenarios. We also measure the ability of the AIC and the BIC criteria to select the correct number of mixture components. Simulations are performed using the R language \citep{Rcoreteam} and C++ with the Rcpp package \citep{Rcpp2011}. Programs are available on request to the authors.
 
\subsection{Simulation protocol and indicators of performance}

In order to get realistic simulations, we simulated qualitative trajectories based on semi-Markov chains whose parameters were estimated  on the real dataset presented in the Introduction of the paper. In that experiment, panelists evaluated three different chocolates, with a list of $D=10$ attributes, the first one with 70\% of cocoa, the  second one with 70\% of cocoa too but sweeter than the first one and a third one  with 90\% of cocoa (see \citep{Visalli2016} for a more detailed presentation of the data). An experiment related to the tasting of the first chocolate is presented in Figure~\ref{fig01}. The components of the renewal Markov process corresponding to each chocolate are estimated by maximum likelihood (see \cite{Lecuelle}), considering gamma distributed sojourn times with no anticipation effect, as in Section~\ref{sec:SMCGamma}. We also evaluate the effect of the introduction of the penalty~(\ref{def:penalty}) on the accuracy of the estimates.

First, we consider a known number of components equal to two, with simulated sequences of 4 or 10 transitions, without any absorbing state. We study two cases of mixtures: the fist one with two well separated sub populations (the chocolates with 70\% and 90\% of cocoa) and the second one with two populations with similar distributions (the two chocolates with 70\% of cocoa).

Second, we assume that the number of components is unknown to evaluate the ability of the different information criteria presented in Section~\ref{sec:BIC} to recover the true number of components in the population. We consider three different configurations. One with only one component (chocolate with 70\% of chocolate), one with two well separated components (the chocolates with 70\% and 90\% of cocoa) and one with two similar components (the two chocolates with 70\% of cocoa). The selection of the number of components is a difficult task and the information criteria do not always give good results with stochastic processes (see for example \citep{Celeux2008}).

Thanks to our knowledge of these chocolates, we can assume that some transitions are not possible (occur with a probability zero). Taking this information into account, we can reduce the number of transition parameters to be estimated. We have 49 unknown probability transition parameters for the chocolate with 70\% of cocoa, 62 unknown parameters when considering the two chocolates with 70\% of cocoa and 69 unknown parameters when considering the chocolate with 70\% of cocoa and the one with 90\% of cocoa.

For simulating mixtures with $G=2$ components, the number of individuals belonging to each component is randomly selected thanks to a binomial law $B(n,0.5)$, meaning that $\pi_1 = \pi_2 = 1/2$. Then, for each type of chocolate, individual trajectories are simulated sequentially by selecting randomly the successive states and durations according to the estimated transition probabilities and dominance duration distributions. For each case, we simulated 500 datasets with sample of sizes $n=60$, $n=200$ and $n=600$ and $B=3$ replications. 
%For each individual 3 sequences corresponding to 3 repetitions of the tasting of one product (as in the dataset presented in results) are simulated.  

In order to avoid computation issues when estimating the parameters related to the gamma distributions, the values of $\widehat{Z}_{ig}^{(m)}$ are rounded to $10^{-4}$ and the maximum likelihood estimation is only performed when there are more than 7 observations. Otherwise the gamma parameters are set to the  values estimated on all the observations belonging to the corresponding mixture, independently of the state.

The number of maximal iterations of the EM algorithm is set to 100. \textit{A posteriori} this was large enough because, for all the considered designs, convergence was achieved before 100 iterations.

To check if the transition matrices are well estimated, we consider the following relative error between the estimated transition matrices $\widehat{\mathbf{P}}^g$ and the transition matrices $\mathbf{P}^g$ used to generate the simulated data for component $g$:
\begin{align}
\mbox{Err} (\mathbf{P}^g) &=  \frac{\left\| \mathbf{P}^g - \widehat{\mathbf{P}}^g \right\|_2^2}{\left\|\mathbf{P}^g\right\|_2^2}, 
\label{def:errP}
\end{align}
where $\left\|\mathbf{P}\right\|_2 = \mbox{tr}\left(\mathbf{P}'\mathbf{P}\right)$ is the squared Frobenius norm of matrix $\mathbf{P}$. 
A similar error is computed for the initial probabilities: 
\begin{align}
\mbox{Err} (\boldsymbol{\alpha}^g) &=  \frac{\left\| \boldsymbol{\alpha}^g - \widehat{\boldsymbol{\alpha}}^g \right\|_2^2}{\left\|\boldsymbol{\alpha}^g\right\|_2^2}. 
\label{def:errAlpha}
\end{align}
We also check if the estimated parameters of the sojourn time gamma distribution are well estimated by  considering the following relative errors
\begin{align}
\mbox{Err}(a) = \frac{\sum_{l=1}^D \sum_{g=1}^G  ((\widehat{a}_{l}^g)-a_{l}^g)^2}{\sum_{l=1}^D \sum_{g=1}^G  (a_{l}^g)^2}, \qquad \mbox{Err}(\lambda) =  \frac{\sum_{l=1}^D \sum_{g=1}^G  ((\widehat{\lambda}_{l}^g)-\lambda_{l}^g)^2}{\sum_{l=1}^D \sum_{g=1}^G  (\lambda_{l}^g)^2}. 
\label{def:erra}
\end{align}

\subsection{Results}

\begin{table}
\caption{\label{simu_sans_penal}Parameter estimation errors when considering unpenalized EM for two clusters  with $n=60, n=200$ and $n=600$ and with simulated sequences with 4 and 10 transitions and $B=3$ repetitions. For each design, mean and standard deviation, in brackets,  are computed considering 500 simulated datasets.}
\centering
\begin{tabular}{lllllllll}
%\noalign{\smallskip} \hline\hline \noalign{\smallskip}
& Parameters & $\mbox{Err} (\boldsymbol{\alpha}^1)$ & $\mbox{Err} (\boldsymbol{\alpha}^2)$ & $\mbox{Err} (\mathbf{P}^1)$ & $\mbox{Err} (\mathbf{P}^2)$ & $\mbox{Err}(a)$ & $\mbox{Err}(\lambda)$ & $\pi_1 =0.5$\\
\hline
\multicolumn{3}{l}{\textbf{With 4 transitions}}\\
\hline
& \multicolumn{3}{l}{Well separated components}\\
%&True values & 0 & 0 & 0 & 0 & 0 & 0 & 0.5 \\
\hline
&$n=60$ & .01(.01) & .01(.02) & .26(.12) & .18(.10) & .23(.75) & .40(.91) & .55(.14)\\
&$n=200$ & <.01(<.01) & <.01(<.01) & .06(.04) & .04(.02) & .04(.06) & .08(.14) & .50(.04)\\
&$n=600$ & <.01(<.01) & <.01(<.01) & .02(.01) & .01(<.01) & .01(.01) & .02(.02) & .50(.02)\\

\hline
&\multicolumn{3}{l}{Not well separated}\\
%&True values & 0 & 0 & 0 & 0 & 0 & 0 & 0.5 \\
\hline
&$n=60$ & .01(.01) & .03(.04) & .33(.13) & .47(.15) & .44(1.77) & .70(3.00) & .61(0.25)\\
&$n=200$ & <.01(<.01) & .01(.03) & .10(.08) & .16(.16) & .29(2.18) & .38(2.93) & .49(.12)\\
&$n=600$ & <.01(<.01) & <.01(<.01) & .02(.01) & .01(.03) & .03(.06) & .04(.12) & .50(.03)\\
\hline
\multicolumn{3}{l}{\textbf{With 10 transitions}}\\
\hline
&\multicolumn{3}{l}{Well separated components}\\
%&True values & 0 & 0 & 0 & 0 & 0 & 0 & 0.5 \\
\hline
&$n=60$ & .01(.01) & .01(.01) & .08(.06) & .05(.04) & .07(.12) & .15(.21) & .50(.10)\\
&$n=200$ & <.01(<.01) & <.01(<.01) & .02(.01) & .01(<.01) & .01(.01) & .02(.02) & .50(.04)\\
&$n=600$ & <.01(<.01) & <.01(<.01) & .01(<.01) & <.01(<.01) & <.01(<.01) & .01(<.01) & .50(.02)\\
\hline
&\multicolumn{3}{l}{Not well separated}\\
%&True values & 0 & 0 & 0 & 0 & 0 & 0 & 0.5 \\
\hline
&$n=60$ & .01(.01) & .03(.07) & .09(.06) & .23(.19) & .18(.41) & .21(.85) & .62(.19)\\
&$n=200$ & <.01(<.01) & <.01(<.01) & .02(.01) & .02(.06) & .03(.04) & .03(.04) & .53(.07)\\
&$n=600$ & <.01(<.01) & <.01(<.01) & .01(<.01) & <.01(<.01) & .01(.01) & .01(.01) & .50(.02)\\
\end{tabular}
\end{table}

{\scriptsize
\begin{table}
\caption{\label{simu_parameters}Parameter estimation errors when considering penalized EM for two clusters  with $n=60, n=200$ and $n=600$ and with simulated sequences with 4 and 10 transitions and $B=3$ repetitions. For each design, the mean and the standard deviation, in brackets,  are computed considering 500 simulated datasets.}
\centering
\begin{tabular}{lllllllll}
%\noalign{\smallskip} \hline\hline \noalign{\smallskip}
& Parameters & $\mbox{Err} (\boldsymbol{\alpha}^1)$ & $\mbox{Err} (\boldsymbol{\alpha}^2)$ & $\mbox{Err} (\mathbf{P}^1)$ & $\mbox{Err} (\mathbf{P}^2)$ & $\mbox{Err}(a)$ & $\mbox{Err}(\lambda)$ & $\pi_1=0.5$\\
\hline
\multicolumn{3}{l}{\textbf{With 4 transitions}}\\
\hline
& \multicolumn{3}{l}{Well separated components}\\
%&True values & 0 & 0 & 0 & 0 & 0 & 0 & 0.5 \\
\hline
&$n=60$ & .01(.01) & .01(.02) & .26(.13) & .18(.10) & .10(.07) & .24(.15) & .54(.15)\\
&$n=200$ & <.01(<.01) & <.01(<.01) & .06(.04) & .04(.02) & .03(.03) & .06(.07) & .50(.05)\\
&$n=600$ & <.01(<.01) & <.01(<.01) & .02(.01) & .01(<.01) & .01(<.01) & .01(.01) & .50(.02)\\

\hline
&\multicolumn{3}{l}{Not well separated}\\
%&True values & 0 & 0 & 0 & 0 & 0 & 0 & 0.5 \\
\hline
&$n=60$ & .01(.01) & .04(.08) & .32(.13) & .48(.16) & .11(.11) & .22(.42) & .61(.26)\\
&$n=200$ & <.01(<.01) & .01(.02) & .10(.07) & .15(.16) & .09(.19) & .11(.22) & .50(.12)\\
&$n=600$ & <.01(<.01) & <.01(<.01) & .02(.01) & .01(.05) & .03(.22) & .04(.30) & .50(.03)\\
\hline
\multicolumn{3}{l}{\textbf{With 10 transitions}}\\
\hline
&\multicolumn{3}{l}{Well separated components}\\
%&True values & 0 & 0 & 0 & 0 & 0 & 0 & 0.5 \\
\hline
&$n=60$ & .01(.01) & .01(.01) & .08(.07) & .06(.05) & .06(.04) & .13(.11) & .49(.11)\\
&$n=200$ & <.01(<.01) & <.01(<.01) & .01(.01) & .01(.01) & .01(.01) & .02(.03) & .50(.04)\\
&$n=600$ & <.01(<.01) & <.01(<.01) & .01(<.01) & <.01(<.01) & <.01(<.01) & .01(<.01) & .50(.02)\\

\hline
&\multicolumn{3}{l}{Not well separated}\\
%&True values & 0 & 0 & 0 & 0 & 0 & 0 & 0.5 \\
\hline
&$n=60$ & .01(.01) & .02(.04) & .10(.06) & .20(.18) & .09(.14) & .10(.20) & .62(.18)\\
&$n=200$ & <.01(<.01) & <.01(<.01) & .02(.01) & .01(.04) & .03(.02) & .03(.03) & .53(.07)\\
&$n=600$ & <.01(<.01) & <.01(<.01) & .01(<.01) & <.01(<.01) & .01(<.01) & .01(<.01) & .50(.02)\\
\end{tabular}
\end{table}
}
Parameter estimation errors, evaluated with (\ref{def:errP}), (\ref{def:errAlpha}) and (\ref{def:erra}), are given in Table~\ref{simu_sans_penal} for the unpenalized version of the EM algorithm and in Table~\ref{simu_parameters} when the penalized version of the EM algorithm described in (\ref{def:mlepen}) is employed to estimate the parameters. 
We note that the introduction of the penalty allows to improve the accuracy of the estimates, especially for small samples, few transitions,  or with clusters with similar distribution of the semi-Markov processes. Without penalty, we observe larger mean errors for the estimated parameters of the sojourn time distribution and high values for the standard deviations of the errors. For example, when  $n=60$ with only 4 observed transitions and clusters that are not well separated, we obtain $Err(\lambda)=0.70$ without penalty whereas this error is reduced to $Err(\lambda)=0.22$ thanks to the introduction of the penalty.
When the sample size gets larger ($n=200$ or $n=600$) and the number of transitions is large both estimation procedures lead to similar results.

From now on, we will only consider estimates obtained with the penalized EM algorithm.

\begin{table}
\caption{\label{classrate}Correct classification rate for two clusters with well separated components and not well separated components with $n=60, n=200$ and $n=600$ and with length of simulated sequences equal to 4 and 10 transitions. For each design, mean and standard deviation, in brackets, are computed from 500 simulated datasets.}
\centering
\begin{tabular}{lllllllll}
\noalign{\smallskip} \hline\hline \noalign{\smallskip}
&& \multicolumn{3}{c}{\textbf{Well separated components}}  & & \multicolumn{3}{c}{\textbf{Not well separated components}} \\
\cline{3-5}
\cline{7-9}
&& $n=60$ & $n=200$ & $n=600$ & & $n=60$ & $n=200$ & $n=600$\\
\hline
\multicolumn{3}{l}{\textbf{With 4 transitions}}\\
\hline
&k-means 
& .85(.07) & .86(.05) & .86(.03) & & .81(.09) & .78(.08) & .76(.06)\\
\hline
&Mixture model 
& .92(.07) & .99(.02) & 1(<.01) & & .82(.09) & .93(.06) & .98(.02)\\
\hline
\multicolumn{3}{l}{\textbf{With 10 transitions}}\\
\hline
&k-means 
& .87(.07) & .89(.04) & .90(.02) & & .83(.07) & .84(.05) & .85(.03)\\
\hline
&Mixture model 
& .97(.05) & 1(.01) & 1(<.01) & & .89(.09) & .97(.05) & 1(<.01)\\
\noalign{\smallskip} \hline\hline \noalign{\smallskip}
\end{tabular}
\end{table}

In our simulation context, we know for each trajectory which component of the mixture it belongs to and we can check if it has been assigned with the MAP criterion to the right component.  The rate of correct classification is given in Table~\ref{classrate}. We note that overall, the rate of well classified trajectories is high with values ranging from $0.83$ to $1$. Model-based clustering substantially improves the classification accuracy compared to k-means, except for the more difficult case with a small sample ($n=60$), 4 transitions and clusters not well separated, were both approaches do not perform well.

\begin{table}
\caption{\label{componentchoice}Choice of the number of components with one component, two well separated components and two not well separated components and 4 or 10 observed transitions. The number of clusters selected by the BIC and the AIC are shown for 500 simulated datasets.}
\centering 
\begin{tabular}{llllllllllllll} 
\noalign{\smallskip} \hline\hline \noalign{\smallskip}
&Selected number & & \multicolumn{3}{c}{One component}  & & & & \multicolumn{4}{c}{Two components}  \\
&of components & & & & & & \multicolumn{3}{c}{Well separated} & & \multicolumn{3}{c}{Not well separated} \\
\cline{4-6}
\cline{8-10}
\cline{12-14}
&& $n$ & $60$ & $200$ & $600$ & & $60$ & $200$ & $600$ & & $60$ & $200$ & $600$\\
\hline
\multicolumn{3}{l}{\textbf{With 4 transitions}}\\
\hline
&BIC\\
\hline
&1 & & 500 & 500 & 500 && 500 & 5 & 0 && 500 & 491 & 0\\
&2 & & 0 & 0 & 0 && 0 & 493 & 500 && 0 & 9 & 494\\
&3 & & 0 & 0 & 0 & & 0 & 2 & 0 && 0 & 0 & 6\\
\hline
&AIC\\
\hline
&1 & & 500 & 500 & 500 && 93 & 0 & 0 && 491 & 4 & 0\\
&2 & & 0 & 0 & 0 && 394 & 473 & 498 && 9 & 373 & 487\\
&3 & & 0 & 0 & 0 && 13 & 27 & 2 && 0 & 123 & 13\\

\hline
\multicolumn{3}{l}{\textbf{With 10 transitions}}\\
\hline
&BIC\\
\hline
&1 & & 500 & 500 & 500 && 43& 0 & 0 && 411 & 0 & 0\\
&2 & & 0 & 0 & 0 && 457 & 497 & 500 && 89 & 431 & 495\\
&3 & & 0 & 0 & 0 & & 0 & 3 & 0 && 0 & 69 & 5\\
\hline
&AIC\\
\hline
&1 & & 500 & 500 & 500 && 0 & 0 & 0 && 27 & 0 & 0\\
&2 & & 0 & 0 & 0 && 407 & 491 & 498 && 245 & 318 & 495\\
&3 & & 0 & 0 & 0 & & 93 & 9 & 2 && 228 & 182 & 15\\
\noalign{\smallskip} \hline\hline \noalign{\smallskip}
\end{tabular}
\end{table}

We present in Table~\ref{componentchoice} the number of components selected by the BIC and the AIC. Whatever the number of individuals, the BIC and the AIC select the correct number of components when there is only one component. With two well separated clusters, the BIC and the AIC generally give good results, except for the case with 4 transitions and $n=60$ where the BIC and, to a lesser degree, the AIC, select only one component rather than two. 

When the two mixture components are not very different, the BIC and the AIC only provide effective criteria for selecting the number of components when the samples are large. The AIC often selects the same number of components as the BIC, but it sometimes selects too many components. For small samples and small number of transitions, the BIC criterion is more restrictive and tends to lead to an underestimation of  the true number of components. Similar conclusions, in a different context, are found in \cite{Celeux2008}. The $\mbox{AIC}_c$ can only be used with large samples because of the too large number of parameters of the model. It does not perform better than the BIC and the AIC in this simulation study and the corresponding results are not shown here.

\section{Clustering Temporal Dominance of Sensations for a Gouda cheese}\label{sec:gouda}

We now study data resulting on an experiment of the European Sensory Network (ESN) aiming at measuring simultaneously perception and liking of Gouda cheeses \citep{Thomas2017}. A large panel of $n=665$ consumers from 6 european countries tasted 4 Gouda cheeses with different ages and fat contents according to the Temporal Dominance of Sensations protocol. A list of $D=10$ attributes was presented to the consumers on a computer screen. Panelists tasted $B=3$ successive bites so there is 3 sequences corresponding to the 3 repetitions for each panelist and for each product. 
In this sample, the mean number of transitions within a sequence is equal to 4.1 (see Table~\ref{GCmeannumber} in the Appendix). 

Our goal in this study is to perform a segmentation of the panel, and to describe, if there are any, the differences of perceptions for a product. We only present the results for a young and low fat Gouda cheese, whose perception by consumers is more complex.

The maximal number of iterations of the EM algorithm is set to 400 because we observed that the algorithm requires more iterations to converge with this dataset. This can be explained by a higher complexity of the model than in the simulation study because all transitions are possible with these products.

\begin{table}
\caption{\label{criterion} Values taken by the BIC, the AIC and the $\mbox{AIC}_c$ for a number of clusters $G$ ranging from 1 to 4 for the young and low fat Gouda cheese from the ESN dataset.}
\begin{tabular}{lcccc}
\noalign{\smallskip} \hline\hline \noalign{\smallskip}
$G$ & 1 & 2 & 3 & 4 \\
\hline
BIC & 87449.96 & 86720.10 & 86830.30 & 87295.43 \\
\hline
AIC & 86839.73 & 85494.05 & 84988.43 & 84837.74 \\
\hline
$\mbox{AIC}_c$ & 86882.94 & 85710.59 & 85636.61 & 86554.71 \\
\noalign{\smallskip} \hline\hline \noalign{\smallskip}
\end{tabular}
\end{table}

As shown in Table~\ref{criterion},  all the information criteria approaches select at least two mixture components, showing the existence of different behaviors in the panel. The BIC suggests to consider two clusters and the $\mbox{AIC}_c$ suggests to consider three clusters but both take really close values for two and three components. That is why, we will examine these two cases in the following. The AIC suggests to consider at least 4 clusters, but as it well known, it is a less parsimonious criterion than the BIC and the $\mbox{AIC}_c$ which generally leads to consider a too large number of mixture components. With two components, the obtained clusters are respectively composed of 398 and 267 individuals, whereas with three components, the obtained clusters are respectively composed of 242, 209 and 214 individuals.

\begin{table}
\caption{\label{intialproba}Estimated initial probabilities for  the young and low fat Gouda cheese from the ESN dataset, considering two or three mixture components.}
\begin{tabular}{lllllllllll}
\noalign{\smallskip} \hline\hline \noalign{\smallskip}
\textit{Cluster} & \multicolumn{10}{c}{\textit{Estimates for the following attributes:}}  \\
\cline{2-11}
& Bitter & Cheese & Dense & Fatty & Melting & Milky & Salty & Sharp & Sour & Tender \\
&  &  & hard &  &  & cream &  &  &  &  \\
\hline
\multicolumn{3}{l}{\textbf{With 2 components}}\\
\hline
1 & .03 & .08 & .44 & .05 & .03 & .04 & .03 & .01 & .02 & .27 \\
2 & .02 & .04 & .39 & .08 & .05 & .05 & .03 & .02 & .01 & .32 \\
\hline
\multicolumn{3}{l}{\textbf{With 3 components}}\\
\hline
1 & .02 & .10 & .15 & .07 & .06 & .07 & .04 & .01 & .02 & .46 \\
2 & .02 & .04 & .41 & .08 & .04 & .04 & .02 & .02 & .01 & .32 \\
3 & .03 & .05 & .75 & .03 & 0 & 0 & .03 & .02 & .02 & .06 \\

\noalign{\smallskip} \hline\hline \noalign{\smallskip}
\end{tabular}
\end{table}

The estimated initial probabilities are shown in Table~\ref{intialproba}. As expected  in sensory studies, most of the panelists choose a texture attribute (Dense hard or Tender) as first dominant attribute. With two components, the initial probabilities are really close with only some small differences. On the other hand, with  three components, large differences are observed between clusters especially for the attributes Dense hard and Tender. In  cluster one, most of the panelists chose Tender as first attribute whereas in cluster three, most of the panelists chose Dense hard. In cluster two, both Dense hard and Tender have a high probability to be chosen as first attribute.

\begin{figure}
\centering
\makebox{\includegraphics[width=0.8 \textwidth]{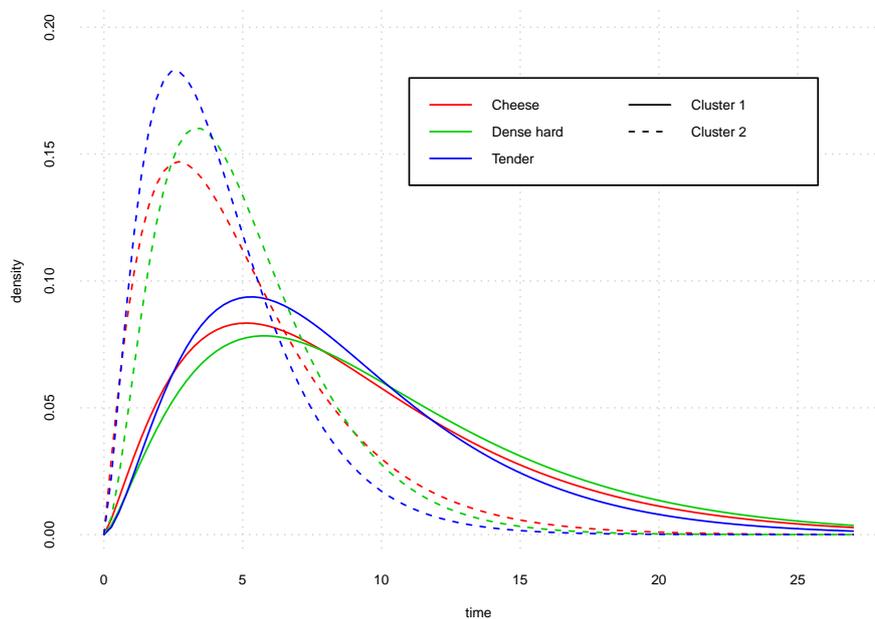}}
\caption{\label{paramgamma2clust}Estimated sojourn time distributions for the attributes Cheese, Dense hard and Tender when considering two  mixture components for  the young and low fat Gouda cheese.}
\end{figure}

\begin{figure}
\centering
\makebox{\includegraphics[width=0.8 \textwidth]{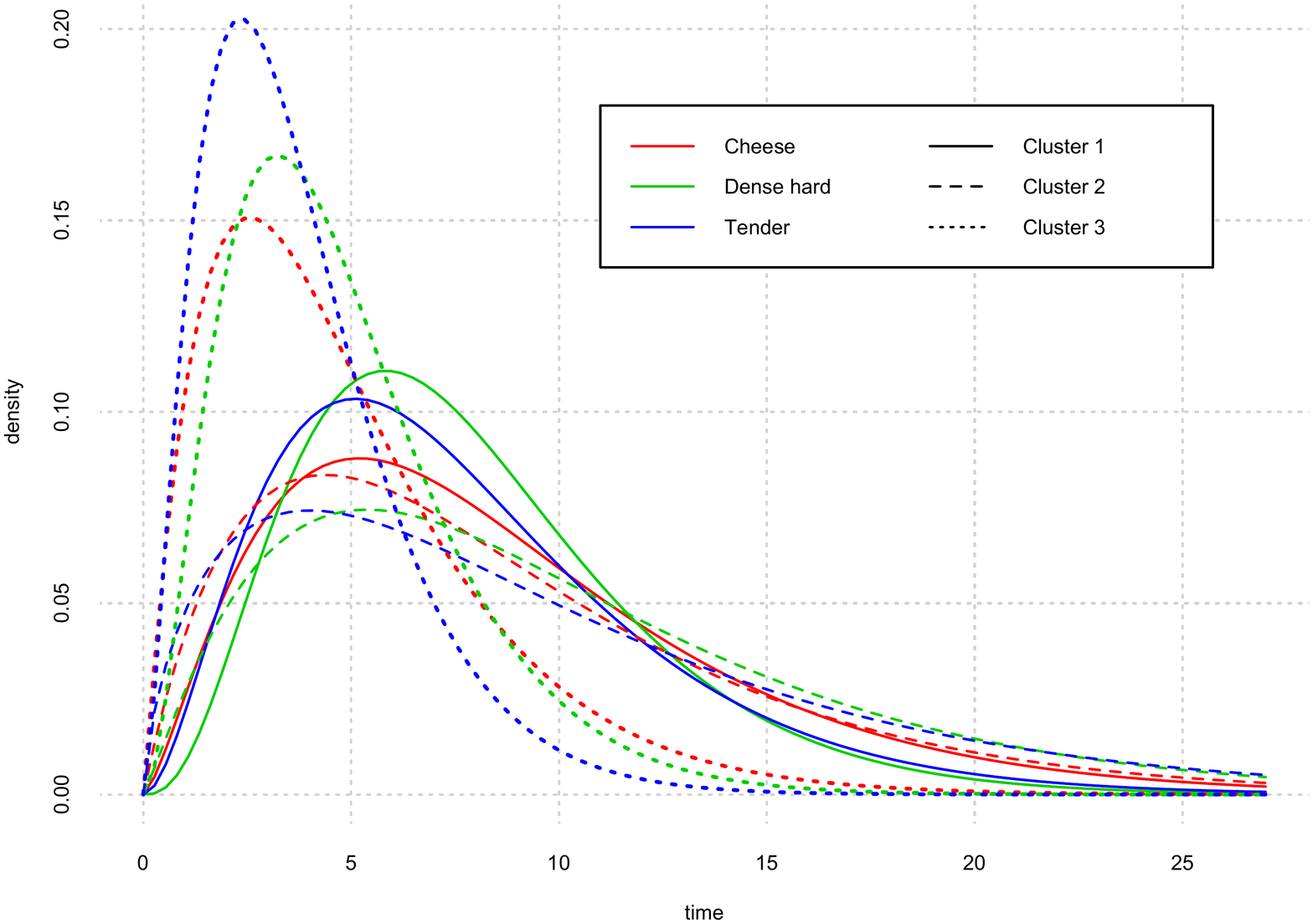}}
\caption{\label{paramgamma3clust}Estimated sojourn time distributions for the attributes Cheese, Dense hard and Tender when considering three components for  the young and low fat Gouda cheese.}
\end{figure}

Figure~\ref{paramgamma2clust} presents the estimated gamma distributions of the sojourn times, with two components, for the attributes Cheese, Dense hard and Tender. Figure~\ref{paramgamma3clust} presents the estimated sojourn time distributions when considering three components. We can note that for all clusters, there are only small differences between the estimated distributions of the different attributes. Then, we can observe that with two components, the estimated distributions are different between the two clusters, with higher probabilities for long durations in cluster one. With three components, the estimated distributions are really similar for the clusters one and three but are different from cluster two. 

% ajouter graphes transition

\begin{figure}
\centering
\makebox{\includegraphics[width=0.5 \textwidth]{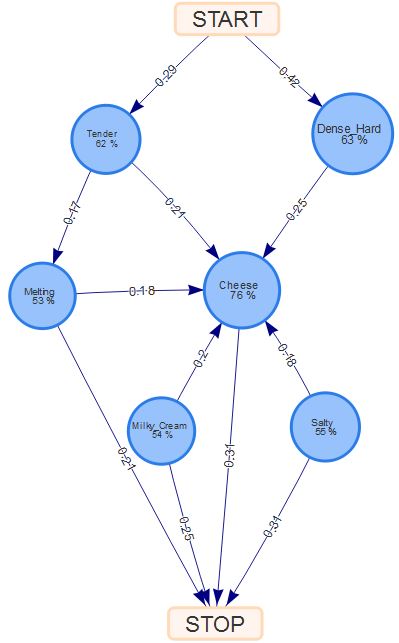}}
\caption{\label{graphGouda}TDS graph for the young and low fat Gouda cheese.}
\end{figure}

\begin{figure}
\begin{tabular}{cc}
 \includegraphics[width=0.4 \textwidth]{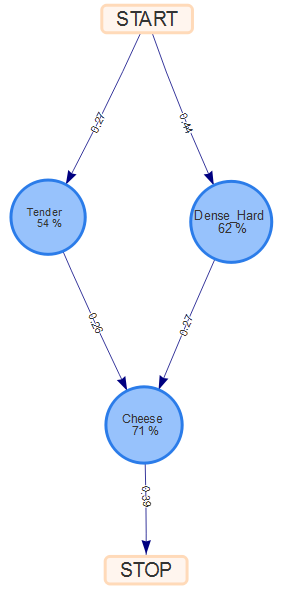} &
 \includegraphics[width=0.6 \textwidth]{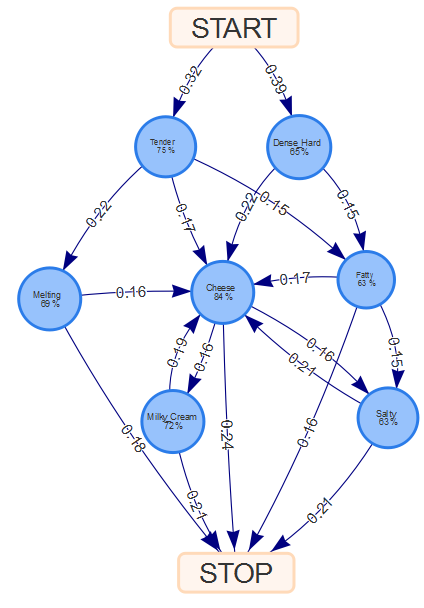}\\
\end{tabular}
\caption{\label{graph2Clust}TDS graphs for the young and low fat Gouda cheese when clustering into 2 segments.}
\end{figure}

\begin{figure}
\begin{tabular}{cc}
 \includegraphics[width=0.48 \textwidth]{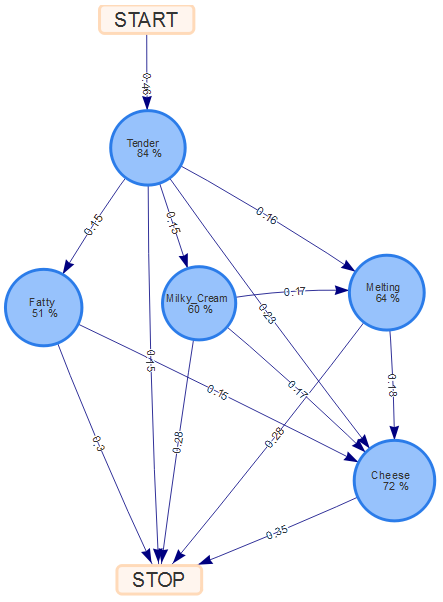} &
 \includegraphics[width=0.52 \textwidth]{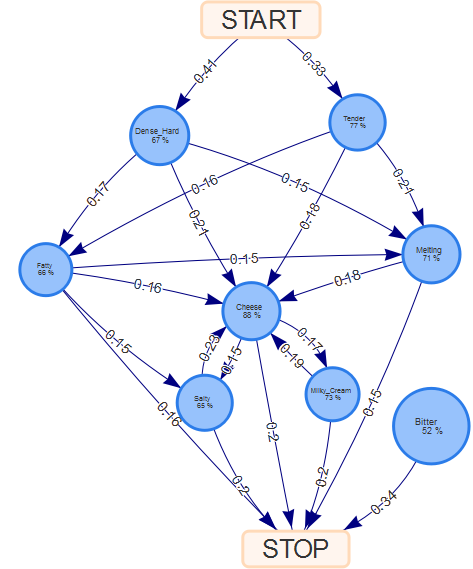}\\
 \multicolumn{2}{c}{\includegraphics[width=0.7 \textwidth]{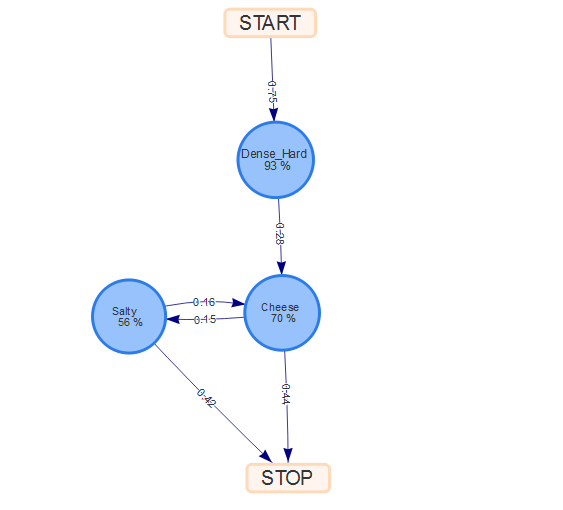}} \\
\end{tabular}
\caption{\label{graph3Clust}TDS graphs for the young and low fat Gouda cheese when clustering into 3 segments.}
\end{figure}

TDS graphs of Figure 5-7 represent the most important transitions, namely those with a probability larger than 0.15 and with at least one half of the panelists having actually elicited the attribute of the  product.
The TDS graph from the whole group (Figure 5) suggests the existence of two different sequences of perception: the first one starts with the attribute "Dense Hard", transits to "Cheese" and ends, whereas the second one starts with "Tender" and either goes to "Cheese", then ends, or goes to "Melting" and then either ends or goes to "Cheese" before ending. It is interesting to note that the "Milky Cream" and "Salty" attributes are present on the graph, since elicited by 54\% and 55 \% of the panelists, but are reached by no arrows, since every transition to them occurred with a probability lower than 0.15.
Considering the segmentation into  two clusters, Figure 6 presents the two TDS graphs associated to each cluster. Both clusters start with a more or less balanced choice between "Dense Hard" and "Tender".  From that point, panelists of Cluster 1 move to "Cheese" and then ends, whereas panelist of cluster 2 followed a more complex route. Indeed, those being on "Tender" can move to "Melting", Cheese" or "Fatty" and those on "Dense Hard" to the same but "Melting". Then their route to the end can be quite complex using some transitions to "Milky Cream" or Salty", the two attributes having too small probabilities at panel level to be reached. The fact that both group starts, as the whole panel, with a choice between opposite attribute "Tender" and "Dense Hard" is not satisfying and claims for investigating the decomposition into three clusters.
Indeed, Clusters 1 and 3 in Figure 7 (segmentation into 3 groups) start respectively with "Tender" and "Dense Hard" and then follow a different route: Cluster going directly to "Cheese", whereas Cluster 1 first transits to "Melting", "Milky Cream" or "Fatty" before reaching the "Cheese" perception. Cluster 1 ends after "Cheese", whereas Cluster 3 can also transit first by "Salty". Cluster 2 in Figure 7 is rather similar to Cluster 2 in Figure 6 exhibiting a very complex perception path. Indeed, Table~\ref{GCmeannumber} shows that panelists in this cluster did an average number of transitions between  5 and 6, whereas panelists in the two other clusters made only an average number of transitions equal to 3. Thus this cluster is likely to gather panelists with different perceptions, but the algorithm was not able to split them while still improving the fit. Therefore, it is also possible that this cluster gathers "noisy panelists", namely panelists having not clearly understood the TDS task.

From a sensory perspective, the segmentation enables to model what is really perceived by the panelists instead of considering  a mean panel overview, corresponding to the perception of none of the panelists. The observed differences between clusters can be explained both by real differences of perception and by differences of behavior with the TDS task. Such mixture models  give the opportunity to further investigate on these new questions by for example examining   the relation between perception and other variables such as age, sex or experience.

\section{Concluding remarks}

This research was motivated by the need of a segmentation method for temporal sensory data. 
For this purpose we have introduced a new mixture of semi-Markov chains which allows, thanks to a model-based clustering approach, to gather into homogeneous groups consumers having similar tasting perceptions.  A penalized EM is introduced to estimate the parameters of the semi Markov chains and the mixture proportions. The evaluation of this estimation method  on simulated data  shows good performance, improving the segmentation obtained by the k-means algorithm, while providing much more information on individual behaviors. The results on real data show an interesting progress in TDS data analysis by offering the possibility of exhibiting different perceptions in a panel for a same product. The development of such segmentation approaches open new perspectives, both for understanding the perception mechanism and for studying how panelists used TDS and understand the TDS protocol.

The models presented in this paper may depend on a large number of parameters and so require to have large samples at hand to be estimated accurately. However, the real data analysis shows that only small differences seem to exist in a same cluster between the gamma distributions modeling sojourn times in the different states. If this hypothesis is verified, we could consider a more parsimonious model by estimating only one gamma distribution for all the states.
As usual with unsupervised classification, choosing the number of clusters is a difficult task. The method used in this paper relies on information criteria and is not very effective for small samples. The BIC criterion seems to overestimate the model complexity whereas AIC has a tendency to select models with a too large complexity.

From a statistical perspective, this sensory modeling issue has given us the opportunity to study a new model for mixtures of qualitative trajectories which may have applications in many fields of science.  The identifiability issue has been addressed under general conditions, considering parametric families of sojourn time distributions.  From a methodological perspective, it also showed  that introducing a penalty in the maximisation step of the EM algorithm improves the quality of the estimates. However some further investigations have to be done to determine whose penalty is the most effective. It would also be of great interest to check rigorously in a future work the consistency of such penalized maximum likelihood approach in the context of mixtures of semi-Markov chains and to study the asymptotic distribution of the estimators. This would permit to build confidence intervals and to test statistical hypotheses.

%%%%%% Bibliographie

%%%###### 
\clearpage

\appendix

\section*{Appendix}

\subsection*{A. Proofs}

\noindent Proof of Proposition 2.1. \\
{\em
The proof is immediate. Consider the  unobserved latent class variable $Z$, taking values in $\{1, \ldots, G\}$ and satisfying $\Pr[Z = g] =\pi_g$ for $g=1, \ldots, G$. The law of $(J_p^\pi,X_p^\pi)_{n\geq 1}$ given $Z=g$ can be expressed as $ \mbox{Law}\left(\boldsymbol{\alpha}^g, \mathbf{P}^g, \Phi_{\ell j}^g, \ell, j \in \mathcal{S} \right)$. Thus, 
\[
\alpha_j^\pi =  \Pr \left[ J_{1}^\pi =j \right]  = \sum_{g=1}^G \Pr \left[ J_{1}^\pi =j | Z=g \right] \Pr [ Z=g]= \sum_{g=1}^G \pi_g \alpha_j^g.
 \] We also clearly have that $(J_p^\pi)_{p \geq 1}$ is a Markov chain, with transition probabilities,
\[
\mathbf{P}_{\ell j}^\pi =  \Pr \left[ J_{p+1}^\pi=j | J_p^\pi =\ell \right] = \sum_{g=1}^G \Pr \left[ J_{p+1}^\pi=j | J_p^\pi =\ell , Z = g \right] \Pr[Z=g] = \sum_{g=1}^G  \pi_g \mathbf{P}_{\ell j}^g,
\]
and, for $t \in T$, 
\[
\Phi_{\ell j}^\pi (t) = \sum_{g=1}^G \Pr \left[X_p \leq t \ | \ J_{p+1}=j, J_{p}=\ell, Z=g \right] \Pr[Z=g] = \sum_{g=1}^G \pi_g \Phi_{\ell j}^g (t).
\]
\hfill $\Box$}

\medskip

\noindent Proof of Proposition 2.2. \\ {\em
We  have, for each mixture component $g$, $\Pr\left[ X_1 \leq t, J_1 = \ell, J_2=j  | Z=g \right] = \alpha_\ell^g  \mathbf{P}_{\ell j}^g \Phi(t, \boldsymbol{\Gamma}^g_{\ell j})$, for $t\geq 0$, $\ell \in \mathcal{S}$ and $j \neq \ell$.  We introduce the $D(D-1)$ dimension (functional) vector, for $t\geq 0$, 
 \begin{align*}
 \mathbf{F}_g(t) &= \left(\alpha_\ell^g  \mathbf{P}_{\ell j}^g \Phi(t, \boldsymbol{\Gamma}^g_{\ell j}), \ell=1,2, \ldots, D, j \neq \ell \right) \\
  & = \left(\alpha_1^g  \mathbf{P}_{1 2}^g \Phi(t, \boldsymbol{\Gamma}^g_{12}),  \cdots,  \alpha_1^g  \mathbf{P}_{1 D}^g \Phi(t, \boldsymbol{\Gamma}^g_{1D}), \cdots, \alpha_D^g  \mathbf{P}_{D (D-1)}^g \Phi(t, \boldsymbol{\Gamma}^g_{D (D-1)}) \right)
\end{align*}
With assumption $\textbf{(H1)}$, all the coefficients $\alpha_\ell^g  \mathbf{P}_{\ell j}^g$ are strictly positive. Since the family of sojourn time distributions is identifiable, we can deduce, with assumption $\textbf{(H2)}$ that $\pi_1 \mathbf{F}_1(t), \pi_2 \mathbf{F}_2(t), \ldots, \pi_g \mathbf{F}_G(t)$ are $G$ linearly independent vectors of functions. Considering the characterization of identifiability established in \cite{YakowitzSpragins1968}, this implies that, up to label swapping, there is a unique way of  writing the mixture distribution $\mathbf{F}^{\pi}(t) = \Pr\left[ X^\pi_1 \leq t, J^\pi_1 = \ell, J^\pi_2=j \right]$, 
%identify, up to label swapping, the unique set of  mixture probabilities $\pi_1, \ldots, \pi_G$ such that
\[
\mathbf{F}^{\pi}(t) = \sum_{g=1}^G \pi_g \mathbf{F}_g(t), \quad t \geq 0.
\]
For $g=1, \ldots, G$, denote by $\mathbf{u}^g(t)=\pi_g \mathbf{F}_g(t)$ the $D(D-1)$ dimensional vector of functions that can be identified from the knowledge of $\mathbf{F}^{\pi}(t)$. Since, by assumption $\textbf{(H1)}$, %the  considered family of sojourn time distributions is identifiable and 
$\pi_g \alpha_\ell^g  \mathbf{P}_{\ell j}^g >0$,  we can determine the value of the set of parameters $\boldsymbol{\Gamma}^g_{\ell j}, \ell=1,2, \ldots, D, j \neq \ell$  by comparing $\pi_g \alpha_\ell^g  \mathbf{P}_{\ell j}^g \Phi(t, \boldsymbol{\Gamma}^g_{\ell j})$  with   $\mathbf{u}_{\ell j}^g(t)$, and we can write  each component of $\mathbf{u}^g(t)$ as follows $u_{\ell j}^g(t)  = \gamma_{\ell j}^g \Phi(t, \boldsymbol{\Gamma}^g_{\ell j})$.  

We prove now that the mixture probability $\pi_g$, the initialization probabilities $\alpha_1^g, \ldots, \alpha_D^g$ and the transition probabilities $\mathbf{P}_{\ell j}^g$ are uniquely determined when the set of coefficients $\{\gamma_{\ell j}^g, \ell \in \mathcal{S}, j\neq \ell\}$ is known. 
Since  $\gamma_{\ell j}^g = \pi_g \alpha_\ell^g  \mathbf{P}_{\ell j}^g$, we first note that 
 \begin{align*}
 \sum_{\ell \in \mathcal{S}} \sum_{j \neq \ell} \gamma_{\ell j}^g & = \pi_g \sum_{\ell \in \mathcal{S}} \sum_{j \neq \ell} \alpha_\ell^g  \mathbf{P}_{\ell j}^g \\
 &= \pi_g
 \end{align*}
 because $\sum_{j \neq \ell} \mathbf{P}_{\ell j}^g =1$ and $\sum_{\ell \in \mathcal{S}} \alpha_\ell^g=1$. Using the same trick again, we get that, for each $\ell \in \mathcal{S}$,
  \begin{align*}
 \frac{1}{\pi_g} \sum_{j \neq \ell} \gamma_{\ell j}^g & =  \sum_{j \neq \ell} \alpha_\ell^g  \mathbf{P}_{\ell j}^g \\
 &= \alpha_\ell^g.
 \end{align*}
Finally, we deduce the values of the transition probabilities, for each $\ell \in \mathcal{S}$ and $j\neq \ell$,
 \begin{align*}
 \mathbf{P}_{\ell j}^g &= \frac{1}{\pi_g  \alpha_\ell^g} \gamma_{\ell j}^g  
\end{align*}
and the proof is complete.
\hfill $\Box$}

%%%%%%%%%%%%%%%%%%%%%%%%%%%%%%%%%%%%%%%%%%%%%%%%%%%

\clearpage
\subsection*{B. Description of the semi-Markov chains used for the simulation study}

\begin{table}
\caption{\label{intialprobaChoc}Estimated initial probabilities for the 3 chocolates.}
\begin{tabular}{l | llllllllll}
\noalign{\smallskip} \hline\hline \noalign{\smallskip}
%\textit{Chocolate} & \multicolumn{10}{c}{\textit{Estimates for the following attributes:}}  \\
%\cline{2-11}
\textbf{Chocolate}& Astringent & Bitter & Cocoa & Crunchy & Dry & Fatty & Melting & Sour & Sweet & Sticky \\
\hline
70 & .00 & .00 & .00 & .81 & .03 & .00 & .03 & .00 & .11 & .03 \\
70 Sweet & .00 & .00 & .00 & .75 & .03 & .00 & .11 & .00 & .06 & .06 \\
90 & .00 & .03 & .03 & .83 & .08 & .00 & .00 & .03 & .00 & .00 \\

\noalign{\smallskip} \hline\hline \noalign{\smallskip}
\end{tabular}
\end{table}

\begin{table}
\caption{\label{transitionsChoc}Estimated transition probabilities for the 3 chocolates.}
\begin{tabular}{lllllllllll}
\noalign{\smallskip} \hline\hline \noalign{\smallskip}
 & \multicolumn{10}{c}{\textbf{Chocolate with 70\% of cocoa}}  \\
\cline{2-11}
& Astringent & Bitter & Cocoa & Crunchy & Dry & Fatty & Melting & Sour & Sweet & Sticky \\
\hline
Astringent & .00 & .33 & .00 & .00 & .00 & .00 & .00 & .33 & .33 & .00 \\
Bitter & .06 & .00 & .25 & .00 & .00 & .06 & .13 & .06 & .31 & .11 \\
Cocoa & .00 & .15 & .00 & .00 & .15 & .09 & .21 & .06 & .27 & .06 \\
Crunchy & .00 & .07 & .40 & .00 & .17 & .00 & .03 & .00 & .27 & .07 \\
Dry & .00 & .15 & .15 & .00 & .00 & .15 & .00 & .15 & .38 & .00 \\
Fatty & .00 & .13 & .38 & .00 & .00 & .00 & .38 & .00 & .13 & .00 \\
Melting & .00 & .00 & .21 & .00 & .00 & .00 & .00 & .11 & .58 & .11 \\
Sour & .09 & .36 & .27 & .00 & .00 & .00 & .18 & .00 & .00 & .09 \\
Sweet & .03 & .03 & .28 & .05 & .03 & .08 & .28 & .10 & .00 & .13 \\
Sticky & .00 & .00 & .00 & .10 & .20 & .00 & .20 & .10 & .40 & .00 \\
\hline \noalign{\smallskip}
 & \multicolumn{10}{c}{\textbf{Chocolate with 70\% of cocoa sweet}}  \\
\cline{2-11}
& Astringent & Bitter & Cocoa & Crunchy & Dry & Fatty & Melting & Sour & Sweet & Sticky \\
\hline
Astringent & .00 & .00 & .00 & .00 & .00 & .00 & 1.00 & .00 & .00 & .00 \\ 
Bitter & .00 & .00 & .00 & .00 & .00 & 1.00 & .00 & .00 & .00 & .00 \\ 
Cocoa & .03 & .03 & .00 & .03 & .00 & .23 & .34 & .00 & .34 & .00 \\ 
Crunchy & .00 & .00 & .23 & .00 & .16 & .03 & .10 & .00 & .45 & .03 \\ 
Dry & .00 & .00 & .29 & .14 & .00 & .29 & .00 & .00 & .14 & .14 \\ 
Fatty & .05 & .00 & .36 & .00 & .05 & .00 & .27 & .00 & .23 & .05 \\ 
Melting & .00 & .00 & .25 & .04 & .00 & .18 & .00 & .00 & .54 & .00 \\ 
Sour & .00 & .00 & .00 & .00 & .00 & .00 & 1.00 & .00 & .00 & .00 \\ 
Sweet & .00 & .00 & .46 & .02 & .00 & .17 & .22 & .02 & .00 & .10 \\ 
Sticky & .00 & .00 & .12 & .00 & .00 & .38 & .12 & .00 & .38 & .00 \\
\hline \noalign{\smallskip}
 & \multicolumn{10}{c}{\textbf{Chocolate with 90\% of cocoa}}  \\
\cline{2-11}
& Astringent & Bitter & Cocoa & Crunchy & Dry & Fatty & Melting & Sour & Sweet & Sticky \\
\hline
Astringent & .00 & .53 & .00 & .00 & .00 & .18 & .00 & .06 & .00 & .24 \\ 
Bitter & .19 & .00 & .30 & .00 & .11 & .14 & .07 & .04 & .09 & .07 \\ 
Cocoa & .00 & .48 & .00 & .03 & .10 & .07 & .17 & .00 & .03 & .10 \\ 
Crunchy & .06 & .29 & .13 & .00 & .32 & .13 & .03 & .00 & .00 & .03 \\ 
Dry & .23 & .55 & .18 & .00 & .00 & .00 & .00 & .05 & .00 & .00 \\ 
Fatty & .17 & .44 & .06 & .00 & .00 & .00 & .22 & .00 & .00 & .11 \\ 
Melting & .14 & .57 & .14 & .00 & .00 & .07 & .00 & .00 & .00 & .07 \\ 
Sour & .20 & .60 & .00 & .00 & .00 & .00 & .00 & .00 & .00 & .20 \\ 
Sweet & .00 & .17 & .50 & .00 & .00 & .17 & .17 & .00 & .00 & .00 \\ 
Sticky & .25 & .50 & .00 & .08 & .00 & .08 & .08 & .00 & .00 & .00 \\ 
\noalign{\smallskip} \hline\hline \noalign{\smallskip}
\end{tabular}
\end{table}

\begin{table}
\caption{\label{distributionsChoc}Estimated gamma distributions for the 3 chocolates.}
\begin{tabular}{lllllllllll}
\noalign{\smallskip} \hline\hline \noalign{\smallskip}
 & \multicolumn{10}{c}{\textbf{Chocolate with 70\% of cocoa}}  \\
\cline{2-11}
& Astringent & Bitter & Cocoa & Crunchy & Dry & Fatty & Melting & Sour & Sweet & Sticky \\
\hline
$a$ & 1.90 & 1.38 & 1.53 & 2.83 & 2.12 & 1.78 & 1.72 & 1.18 & 1.73 & 3.45 \\ 
$\lambda$ & 0.29 & 0.21 & 0.21 & 0.41 & 0.38 & 0.21 & 0.35 & 0.15 & 0.26 & 0.77 \\ 
\hline \noalign{\smallskip}
 & \multicolumn{10}{c}{\textbf{Chocolate with 70\% of cocoa sweet}}  \\
\cline{2-11}
& Astringent & Bitter & Cocoa & Crunchy & Dry & Fatty & Melting & Sour & Sweet & Sticky \\
\hline
$a$ & 1.76 & 1.69 & 1.30 & 2.04 & 1.87 & 1.50 & 1.51 & 1.69 & 2.28 & 3.51 \\ 
$\lambda$ & 0.14 & 0.25 & 0.20 & 0.32 & 0.33 & 0.22 & 0.22 & 0.25 & 0.31 & 0.62 \\ 
\hline \noalign{\smallskip}
 & \multicolumn{10}{c}{\textbf{Chocolate with 90\% of cocoa}}  \\
\cline{2-11}
& Astringent & Bitter & Cocoa & Crunchy & Dry & Fatty & Melting & Sour & Sweet & Sticky \\
\hline
$a$ & 1.86 & 1.52 & 1.67 & 2.40 & 2.05 & 3.29 & 2.88 & 1.73 & 3.86 & 3.70 \\ 
$\lambda$ & 0.20 & 0.20 & 0.27 & 0.50 & 0.27 & 0.81 & 0.70 & 0.21 & 1.45 & 0.63 \\ 
\noalign{\smallskip} \hline\hline \noalign{\smallskip}
\end{tabular}
\end{table}

\clearpage
%%%%%%%%%%%%%%%%%%%%%%%%%%%%%%%%%%%%%%%%%%%%%%%%%%%
\subsection*{C. Additional information on the Gouda cheese example}

\begin{table}
\caption{\label{transitionsCheese2Clust} Gouda cheese example: estimated transition probabilities with 2 clusters.}
\begin{tabular}{llllllllllll}
\noalign{\smallskip} \hline\hline \noalign{\smallskip}
 & \multicolumn{11}{c}{\textbf{Cluster 1}}  \\
\cline{2-12}
& Bitter & Cheese & Dense & Fatty & Melting & Milky & Salty & Sharp & Sour & Tender & STOP \\
&  &  & hard &  &  & cream &  &  &  & & \\
\hline
Bitter & 0.00 & 0.08 & 0.05 & 0.05 & 0.05 & 0.04 & 0.08 & 0.04 & 0.06 & 0.03 & 0.53 \\ 
Cheese & 0.07 & 0.00 & 0.05 & 0.07 & 0.09 & 0.08 & 0.13 & 0.04 & 0.04 & 0.05 & 0.39 \\ 
Dense & 0.09 & 0.26 & 0.00 & 0.07 & 0.06 & 0.09 & 0.11 & 0.05 & 0.08 & 0.04 & 0.13 \\ 
Fatty & 0.07 & 0.16 & 0.04 & 0.00 & 0.11 & 0.10 & 0.07 & 0.05 & 0.06 & 0.07 & 0.27 \\ 
Melting & 0.04 & 0.20 & 0.00 & 0.09 & 0.00 & 0.14 & 0.13 & 0.02 & 0.06 & 0.06 & 0.26 \\ 
Milky & 0.04 & 0.20 & 0.01 & 0.07 & 0.13 & 0.00 & 0.11 & 0.02 & 0.04 & 0.08 & 0.30 \\ 
Salty & 0.09 & 0.14 & 0.03 & 0.07 & 0.05 & 0.05 & 0.00 & 0.05 & 0.07 & 0.02 & 0.42 \\ 
Sharp & 0.09 & 0.13 & 0.06 & 0.09 & 0.05 & 0.01 & 0.13 & 0.00 & 0.07 & 0.03 & 0.34 \\ 
Sour & 0.12 & 0.09 & 0.06 & 0.04 & 0.05 & 0.03 & 0.11 & 0.06 & 0.00 & 0.03 & 0.41 \\ 
Tender & 0.04 & 0.26 & 0.02 & 0.15 & 0.12 & 0.13 & 0.09 & 0.03 & 0.03 & 0.00 & 0.13 \\ 
\hline \noalign{\smallskip}
 & \multicolumn{11}{c}{\textbf{Cluster 2}}  \\
\cline{2-12}
& Bitter & Cheese & Dense & Fatty & Melting & Milky & Salty & Sharp & Sour & Tender & STOP \\
&  &  & hard &  &  & cream &  &  &  & & \\
\hline
Bitter & 0.00 & 0.13 & 0.02 & 0.08 & 0.07 & 0.10 & 0.12 & 0.06 & 0.06 & 0.03 & 0.34 \\ 
Cheese & 0.10 & 0.00 & 0.02 & 0.06 & 0.10 & 0.16 & 0.16 & 0.04 & 0.06 & 0.06 & 0.24 \\ 
Dense & 0.07 & 0.23 & 0.00 & 0.15 & 0.12 & 0.08 & 0.10 & 0.07 & 0.04 & 0.10 & 0.04 \\ 
Fatty & 0.05 & 0.17 & 0.04 & 0.00 & 0.14 & 0.13 & 0.14 & 0.03 & 0.05 & 0.09 & 0.16 \\ 
Melting & 0.09 & 0.17 & 0.04 & 0.07 & 0.00 & 0.13 & 0.11 & 0.05 & 0.03 & 0.12 & 0.18 \\ 
Milky & 0.07 & 0.19 & 0.03 & 0.09 & 0.11 & 0.00 & 0.13 & 0.03 & 0.06 & 0.08 & 0.21 \\ 
Salty & 0.13 & 0.21 & 0.03 & 0.06 & 0.08 & 0.09 & 0.00 & 0.05 & 0.09 & 0.05 & 0.22 \\ 
Sharp & 0.14 & 0.12 & 0.05 & 0.09 & 0.11 & 0.09 & 0.14 & 0.00 & 0.04 & 0.05 & 0.17 \\ 
Sour & 0.11 & 0.15 & 0.01 & 0.04 & 0.07 & 0.11 & 0.13 & 0.06 & 0.00 & 0.04 & 0.28 \\ 
Tender & 0.03 & 0.17 & 0.05 & 0.15 & 0.22 & 0.14 & 0.06 & 0.04 & 0.05 & 0.00 & 0.10 \\ 
\noalign{\smallskip} \hline\hline \noalign{\smallskip}
\end{tabular}
\end{table}

\begin{table}
\caption{\label{distributionsCheese2Clust} Gouda cheese example: estimated gamma distributions with 2 clusters.}
\begin{tabular}{lllllllllll}
\noalign{\smallskip} \hline\hline \noalign{\smallskip}
 & \multicolumn{10}{c}{\textbf{Cluster 1}}  \\
\cline{2-11}
& Bitter & Cheese & Dense & Fatty & Melting & Milky & Salty & Sharp & Sour & Tender \\
&  &  & hard &  &  & cream &  &  &  & \\
\hline
$a$ & 1.91 & 2.30 & 2.28 & 2.19 & 2.53 & 2.57 & 2.24 & 2.19 & 2.17 & 2.57 \\ 
$\lambda$ & 0.20 & 0.25 & 0.24 & 0.25 & 0.29 & 0.32 & 0.25 & 0.23 & 0.26 & 0.30 \\ 
\hline \noalign{\smallskip}
 & \multicolumn{10}{c}{\textbf{Cluster 2}}  \\
\cline{2-11}
& Bitter & Cheese & Dense & Fatty & Melting & Milky & Salty & Sharp & Sour & Tender \\
&  &  & hard &  &  & cream &  &  &  & \\
\hline
$a$ & 2.44 & 2.17 & 3.17 & 2.77 & 2.36 & 2.23 & 2.46 & 3.42 & 3.08 & 2.59 \\
$\lambda$ & 0.46 & 0.43 & 0.61 & 0.63 & 0.46 & 0.45 & 0.52 & 0.73 & 0.65 & 0.59 \\ 
\noalign{\smallskip} \hline\hline \noalign{\smallskip}
\end{tabular}
\end{table}

\begin{table}
\caption{\label{transitionsCheese3Clust} Gouda cheese example: estimated transition probabilities with 3 clusters.}
\begin{tabular}{llllllllllll}
\noalign{\smallskip} \hline\hline \noalign{\smallskip}
 & \multicolumn{11}{c}{\textbf{Cluster 1}}  \\
\cline{2-12}
& Bitter & Cheese & Dense & Fatty & Melting & Milky & Salty & Sharp & Sour & Tender & STOP \\
&  &  & hard &  &  & cream &  &  &  & & \\
\hline
Bitter & 0.00 & 0.07 & 0.02 & 0.06 & 0.10 & 0.11 & 0.00 & 0.05 & 0.04 & 0.08 & 0.47 \\ 
Cheese & 0.05 & 0.00 & 0.01 & 0.10 & 0.13 & 0.09 & 0.13 & 0.02 & 0.01 & 0.10 & 0.35 \\ 
Dense & 0.05 & 0.25 & 0.00 & 0.12 & 0.12 & 0.09 & 0.05 & 0.04 & 0.01 & 0.12 & 0.15 \\ 
Fatty & 0.04 & 0.15 & 0.00 & 0.00 & 0.13 & 0.12 & 0.07 & 0.05 & 0.05 & 0.10 & 0.29 \\ 
Melting & 0.05 & 0.19 & 0.01 & 0.10 & 0.00 & 0.14 & 0.08 & 0.03 & 0.04 & 0.09 & 0.28 \\ 
Milky & 0.01 & 0.17 & 0.01 & 0.07 & 0.17 & 0.00 & 0.10 & 0.03 & 0.04 & 0.11 & 0.28 \\ 
Salty & 0.07 & 0.11 & 0.03 & 0.07 & 0.09 & 0.10 & 0.00 & 0.03 & 0.05 & 0.06 & 0.39 \\ 
Sharp & 0.06 & 0.17 & 0.00 & 0.11 & 0.10 & 0.04 & 0.13 & 0.00 & 0.02 & 0.06 & 0.32 \\ 
Sour & 0.11 & 0.04 & 0.00 & 0.10 & 0.09 & 0.05 & 0.05 & 0.01 & 0.00 & 0.04 & 0.50 \\ 
Tender & 0.02 & 0.23 & 0.01 & 0.15 & 0.16 & 0.16 & 0.07 & 0.03 & 0.02 & 0.00 & 0.15 \\ 
\hline \noalign{\smallskip}
 & \multicolumn{11}{c}{\textbf{Cluster 2}}  \\
\cline{2-12}
& Bitter & Cheese & Dense & Fatty & Melting & Milky & Salty & Sharp & Sour & Tender & STOP \\
&  &  & hard &  &  & cream &  &  &  & & \\
\hline
Bitter & 0.00 & 0.13 & 0.01 & 0.08 & 0.07 & 0.10 & 0.12 & 0.06 & 0.06 & 0.02 & 0.34 \\ 
Cheese & 0.10 & 0.00 & 0.02 & 0.06 & 0.11 & 0.17 & 0.16 & 0.04 & 0.07 & 0.06 & 0.21 \\ 
Dense & 0.05 & 0.22 & 0.00 & 0.16 & 0.14 & 0.08 & 0.10 & 0.07 & 0.05 & 0.11 & 0.02 \\ 
Fatty & 0.05 & 0.17 & 0.04 & 0.00 & 0.14 & 0.13 & 0.15 & 0.03 & 0.05 & 0.08 & 0.16 \\ 
Melting & 0.09 & 0.18 & 0.04 & 0.06 & 0.00 & 0.13 & 0.13 & 0.05 & 0.04 & 0.12 & 0.15 \\ 
Milky & 0.08 & 0.19 & 0.02 & 0.10 & 0.10 & 0.00 & 0.13 & 0.03 & 0.07 & 0.08 & 0.20 \\ 
Salty & 0.14 & 0.23 & 0.02 & 0.07 & 0.07 & 0.09 & 0.00 & 0.05 & 0.09 & 0.04 & 0.21 \\ 
Sharp & 0.15 & 0.11 & 0.05 & 0.10 & 0.11 & 0.08 & 0.11 & 0.00 & 0.05 & 0.05 & 0.18 \\ 
Sour & 0.12 & 0.15 & 0.01 & 0.04 & 0.07 & 0.11 & 0.14 & 0.06 & 0.00 & 0.04 & 0.25 \\ 
Tender & 0.03 & 0.18 & 0.05 & 0.16 & 0.21 & 0.12 & 0.07 & 0.04 & 0.05 & 0.00 & 0.09 \\ 
\hline \noalign{\smallskip}
 & \multicolumn{11}{c}{\textbf{Cluster 3}}  \\
\cline{2-12}
& Bitter & Cheese & Dense & Fatty & Melting & Milky & Salty & Sharp & Sour & Tender & STOP \\
&  &  & hard &  &  & cream &  &  &  & & \\
\hline
Bitter & 0.00 & 0.09 & 0.07 & 0.04 & 0.02 & 0.00 & 0.15 & 0.03 & 0.07 & 0.00 & 0.52 \\ 
Cheese & 0.09 & 0.00 & 0.09 & 0.04 & 0.03 & 0.07 & 0.14 & 0.05 & 0.06 & 0.00 & 0.43 \\ 
Dense & 0.11 & 0.27 & 0.00 & 0.06 & 0.04 & 0.09 & 0.12 & 0.06 & 0.09 & 0.02 & 0.13 \\ 
Fatty & 0.12 & 0.21 & 0.14 & 0.00 & 0.08 & 0.06 & 0.08 & 0.02 & 0.09 & 0.03 & 0.19 \\ 
Melting & 0.04 & 0.17 & 0.00 & 0.04 & 0.00 & 0.13 & 0.27 & 0.00 & 0.11 & 0.03 & 0.22 \\ 
Milky & 0.08 & 0.27 & 0.03 & 0.04 & 0.06 & 0.00 & 0.12 & 0.00 & 0.04 & 0.01 & 0.35 \\ 
Salty & 0.11 & 0.16 & 0.05 & 0.06 & 0.03 & 0.02 & 0.00 & 0.07 & 0.08 & 0.01 & 0.42 \\ 
Sharp & 0.10 & 0.10 & 0.11 & 0.06 & 0.02 & 0.02 & 0.18 & 0.00 & 0.10 & 0.01 & 0.30 \\ 
Sour & 0.10 & 0.13 & 0.09 & 0.00 & 0.02 & 0.03 & 0.14 & 0.08 & 0.00 & 0.02 & 0.39 \\ 
Tender & 0.13 & 0.27 & 0.10 & 0.08 & 0.00 & 0.09 & 0.16 & 0.00 & 0.09 & 0.00 & 0.07 \\ 
\noalign{\smallskip} \hline\hline \noalign{\smallskip}
\end{tabular}
\end{table}

\begin{table}
\caption{\label{distributionsCheese3Clust}Gouda cheese example: estimated gamma distributions with 3 clusters.}
\begin{tabular}{lllllllllll}
\noalign{\smallskip} \hline\hline \noalign{\smallskip}
 & \multicolumn{10}{c}{\textbf{Cluster 1}}  \\
\cline{2-11}
& Bitter & Cheese & Dense & Fatty & Melting & Milky & Salty & Sharp & Sour & Tender \\
&  &  & hard &  &  & cream &  &  &  & \\
\hline
$a$ & 2.80 & 2.44 & 2.76 & 2.41 & 2.41 & 2.54 & 3.51 & 2.25 & 2.67 & 2.67 \\ 
$\lambda$ & 0.36 & 0.28 & 0.32 & 0.28 & 0.29 & 0.32 & 0.44 & 0.26 & 0.33 & 0.33 \\ 
\hline \noalign{\smallskip}
 & \multicolumn{10}{c}{\textbf{Cluster 2}}  \\
\cline{2-11}
& Bitter & Cheese & Dense & Fatty & Melting & Milky & Salty & Sharp & Sour & Tender \\
&  &  & hard &  &  & cream &  &  &  & \\
\hline
$a$ & 2.43 & 2.22 & 3.22 & 2.91 & 2.40 & 2.19 & 2.52 & 3.35 & 3.17 & 2.59 \\ 
$\lambda$ & 0.45 & 0.46 & 0.63 & 0.69 & 0.49 & 0.47 & 0.55 & 0.74 & 0.68 & 0.62 \\ 
\hline \noalign{\smallskip}
 & \multicolumn{10}{c}{\textbf{Cluster 3}}  \\
\cline{2-11}
& Bitter & Cheese & Dense & Fatty & Melting & Milky & Salty & Sharp & Sour & Tender \\
&  &  & hard &  &  & cream &  &  &  & \\
\hline
$a$ & 1.55 & 2.07 & 2.11 & 1.67 & 2.75 & 2.87 & 1.76 & 2.24 & 1.96 & 1.58 \\ 
$\lambda$ & 0.16 & 0.23 & 0.23 & 0.21 & 0.32 & 0.39 & 0.20 & 0.23 & 0.24 & 0.20 \\ 
\noalign{\smallskip} \hline\hline \noalign{\smallskip}
\end{tabular}
\end{table}

\begin{table}
\caption{\label{GCmeannumber}Gouda cheese example: mean number of transitions within each cluster for the different scenarios.}
\begin{tabular}{l | lll}
 	& Cluster 1& Cluster 2& Cluster 3 \\ \hline
No segmentation & 	4.13	 & & \\	 
2 clusters	&3.30	& 5.37 &	\\ 
3 clusters	&3.46	& 5.79&	3.26\\
\end{tabular}
\end{table}

\end{document}